\documentclass[sigconf]{acmart}

\usepackage{tabularx}
\usepackage{tcolorbox}
\usepackage{enumitem}
\usepackage{algorithmic}
\usepackage{graphicx}
\usepackage{textcomp}
\usepackage{booktabs} 
\usepackage{threeparttable}
\usepackage{framed}
\usepackage{verbatim}
\usepackage{pifont}
\usepackage{listings}
\usepackage[normalem]{ulem}
\usepackage{xurl} 
\usepackage{tablefootnote}
\usepackage{xcolor}
\usepackage{hyperref}
\hypersetup{
    colorlinks,
    linkcolor={red!50!black},
    citecolor={blue!50!black},
    urlcolor={blue!80!black}
}
\usepackage{filecontents}
\def\BibTeX{{\rm B\kern-.05em{\sc i\kern-.025em b}\kern-.08em
    T\kern-.1667em\lower.7ex\hbox{E}\kern-.125emX}}

\usepackage{color, colortbl}
\usepackage{soul}
\usepackage{multirow}
\usepackage{subcaption}

\DeclareFixedFont{\ttb}{T1}{txtt}{bx}{n}{11} 
\DeclareFixedFont{\ttm}{T1}{txtt}{m}{n}{11}  

\definecolor{deepblue}{rgb}{0,0,0.5}
\definecolor{deepred}{rgb}{0.6,0,0}
\definecolor{deepgreen}{rgb}{0,0.5,0}

\newcommand\pythoninline[1]{{\pythonstyle\lstinline!#1!}}

\definecolor{dkgreen}{rgb}{0,0.6,0}
\definecolor{gray}{rgb}{0.5,0.5,0.5}
\definecolor{GrayMed}{rgb}{0.7,0.7,0.7}
\definecolor{GrayLt}{rgb}{0.9,0.9,0.9}
\definecolor{palegreen}{rgb}{0.92, 1, 0.9}
\definecolor{paleblue}{rgb}{0.92, 0.90, 1}

\definecolor{mauve}{rgb}{0.58,0,0.82}
\acmConference[ICSE 2024]{46th International Conference on Software Engineering}{April 2024}{Lisbon, Portugal}

\AtBeginDocument{%
  \providecommand\BibTeX{{%
    \normalfont B\kern-0.5em{\scshape i\kern-0.25em b}\kern-0.8em\TeX}}}

\copyrightyear{2024}
\acmYear{2024}
\setcopyright{acmlicensed}\acmConference[ICSE '24]{2024 IEEE/ACM 46th International Conference on Software Engineering}{April 14--20, 2024}{Lisbon, Portugal}
\acmBooktitle{2024 IEEE/ACM 46th International Conference on Software Engineering (ICSE '24), April 14--20, 2024, Lisbon, Portugal} 
\acmISBN{979-8-4007-0217-4/24/04}

\begin{document}

\title{High Expectations: An Observational Study of \\ Programming
and Cannabis Intoxication} 

\author{Wenxin He}
\email{wenxinhe@umich.edu}
\affiliation{%
  \institution{University of Michigan}
  \city{Ann Arbor}
  \state{Michigan}
  \country{USA}
}

\author{Manasvi Parikh}
\email{manasvi@umich.edu}
\affiliation{%
  \institution{University of Michigan}
  \city{Ann Arbor}
  \state{Michigan}
  \country{USA}
}

\author{Westley Weimer}
\email{weimerw@umich.edu}
\affiliation{%
  \institution{University of Michigan}
  \city{Ann Arbor}
  \state{Michigan}
  \country{USA}
}

\author{Madeline Endres}
\email{endremad@umich.edu}
\affiliation{%
  \institution{University of Michigan}
  \city{Ann Arbor}
  \state{Michigan}
  \country{USA}
}

\renewcommand{\shortauthors}{Trovato and Tobin, et al.}


\begin{abstract}
Anecdotal evidence of cannabis use by professional programmers abounds. Recent studies have found that some professionals regularly use cannabis while programming, even for work-related tasks. However, accounts of the impacts of cannabis on programming vary widely and are often contradictory. For example, some programmers claim that it impairs their ability to generate correct solutions, while others claim it enhances creativity and focus. 
There remains a need for an empirical understanding of the true impacts of cannabis on programming. This paper presents the first controlled observational study of cannabis's effects on programming ability. Based on a within-subjects design with over 70 participants, we find that, at ecologically valid dosages, cannabis significantly impairs programming performance. \textbf{Programs implemented while high contain more bugs and take longer to write ($p<0.05$)} --- a
small to medium effect ($0.22 \leq d \leq 0.44$).
We also did not find any evidence that high programmers generate more divergent solutions. However, programmers can accurately assess differences in their programming performance ($r=0.59$), even when under the influence of cannabis. 
We hope that this research will facilitate evidence-based policies and help developers make informed decisions regarding cannabis use while programming.
\end{abstract}

\ccsdesc[100]{Social and professional topics~User characteristics}
\ccsdesc[300]{Social and professional topics~Governmental regulations}
\ccsdesc[100]{Social and professional topics~Testing, certification and licensing}
\ccsdesc[100]{Social and professional topics~Employment issues}
\ccsdesc[300]{Human-centered computing~User studies}
\ccsdesc[300]{Software and its engineering~Software development techniques}

\keywords{programming preferences, cannabis, controlled user study, drug policy, preregistered hypotheses}

\if False
\begin{teaserfigure}
  \includegraphics[width=\textwidth]{sampleteaser}
  \caption{Seattle Mariners at Spring Training, 2010.}
  \Description{Enjoying the baseball game from the third-base
  seats. Ichiro Suzuki preparing to bat.}
  \label{fig:teaser}
\end{teaserfigure}

\fi


\maketitle

\section{Introduction}


Software developers commonly use psychoactive
substances while programming~\cite{Endres2022HashingIt}, a behavior at odds
with many company drug and hiring policies~\cite{Newman2023FromOrganizations}. 
\textit{Cannabis sativa} (or ``cannabis'') is the world's most common illicit substance, used by more than 192 million people in 2018~\cite{UNPressReport} and representing a market of 20.5 billion USD~\cite{businessWire}. 
Claims, biases, and folk wisdom about the actual effects of cannabis intoxication on programming do not agree. 
This lack of a firm understanding prevents individuals and companies alike from making informed policy decisions and accurately balancing risk and reward. 

A 2022 study of 800 programmers (including 450 full-time developers) found that 35\% had used cannabis while programming, and
that 18\% did so at least once per month~\cite{Endres2022HashingIt}. Motivations included 
enjoyment, thinking of more creative programming
solutions, enhancing brainstorming, and improving focus~\cite[Tab.~3]{Endres2022HashingIt}. A 2023
study interviewing 25 professional programmers who used psychoactive substances reported 
positive views on the impact of cannabis on brainstorming, neutral views on coding and testing, and negative
views on debugging, design, and documentation~\cite[Tab.~III]{Newman2023FromOrganizations}. 
Psychoactive substances including cannabis are seen as part of the historical tradition of software~\cite{Markoff2005WhatThe}, 
with advocates touting 
focus~\cite{programmingInsider}
and creativity~\cite{simpleProgrammer}
benefits~\cite{Walsh2011DrugsThe}.

In broader contexts, views may be more negative. Cannabis intoxication can impair decision-making accuracy and consistency, resulting in losses
being under-estimated (``treating each loss as a constant and minor negative outcome regardless of the size of the loss'')~\cite{Fridberg2010CognitiveMechanisms}. It can also impair motor control and reaction times. For example, ``acute cannabis intoxication is associated with a statistically significant increase in motor vehicle crash risk''~\cite{Rogeberg2016Cannabis}. 
These more-negative views often inform both general and software-specific regulation. Anti-cannabis hiring and retention policies are prevalent in software companies 
(e.g.,~\cite[pp.~12--13]{ibmHandbook} and~\cite[p.~12]{ciscoHandbook}) and 
almost one-third of developers reported taking a drug test for a programming
job~\cite[Sec.~6]{Endres2022HashingIt}. However, questions have been raised about the efficacy 
of such policies~\cite[Sec.~IX.B]{Newman2023FromOrganizations} and whether they are either needed or beneficial~\cite{bbcFBI2014} in a modern context.

Understanding the impacts of cannabis intoxication on particular aspects of software development
would help fill this knowledge gap,
allowing for evidence-based corporate policies and informed decisions by developers regarding when and if
to use cannabis while programming. An effective understanding would be based on an indicative
sample size,
ecologically-valid conditions 
and both quantitative 
and qualitative 
aspects of produced software. In addition,
any such study must be conducted ethically, given the rapidly-changing legal landscape around cannabis use. 
 
We conducted the first rigorous observational study of the effects of cannabis intoxication on programming. 
We assessed $n=74$ participants from multiple American metropolitan areas across four states and used \emph{pre-registered hypotheses} to mitigate researcher bias. 
Each participant completed two sessions on different days, one while sober and one while using cannabis, in a \emph{randomly-assigned} order. This design permits both within-subject and between-subjects comparisons. 
In each session, participants completed both short targeted programming tasks as well as multiple LeetCode~\cite{Behroozi2019Hiring} 
problems using a standard
development environment (Visual Studio) on their personal computers. In the cannabis
condition, participants were asked to use the amount they would normally use while 
programming, an \emph{ecologically-valid} context that allows us to learn actionable insights. 
Our contributions are:
\begin{enumerate}[leftmargin=15pt,topsep=2pt]
\item{The first rigorous observational study of programming in both sober and cannabis-intoxicated conditions.}
\item{
We find that ecologically-valid cannabis intoxication has a \textbf{small to medium effect on program correctness}: 
programs written by cannabis-intoxicated programmers exhibit \emph{more bugs}, failing 10\% more test cases, on average ($p<0.05$). 
} 
\item{Cannabis-intoxicated programmers take \emph{more time} to write non-trivial functions than do sober programmers (11\% more on time average, $p=0.39$). 
}
\item{Cannabis-intoxicated programmers exhibit \emph{different typing patterns}, including deleting and rewriting code more frequently and pausing for longer without typing ($p \leq 0.003$, $d \ge 0.35$). 
}

\item {
    Despite anecdotes of cannabis improving creativity, we observe no evidence that cannabis-intoxicated programmers make different algorithmic or stylistic programming choices. 
}
\item {
  Cannabis-using programmers accurately recognize their programming performance, even when intoxicated, $r=0.59$.
}
\end{enumerate}

The low impact of cannabis compared to individual
differences, and the ability of developers to recognize cannabis impairment, suggest
caution when crafting anti-cannabis policies (see Section~\ref{sec:discussion}). 

To the best of our knowledge, this is the first study of how of cannabis intoxication
impacts programming. We believe aspects of our design (pre-registered hypotheses, 
ecologically-valid settings, broad participant pool, etc.) make it generalizable and useful 
for informing policies and individual developer decisions surrounding cannabis. We make our (de-identified) replication package available, including raw data, stimuli, analysis scripts, and design documents.

\section{Background and Related Work}
\label{sec:background}

In this section we briefly present related work on cannabis intoxication,
computing-related cannabis use, and software creativity. 

\textbf{Impacts of Cannabis Intoxication}.
Cannabis is well-known for its mind-altering effects. In particular, acute cannabis use (especially for non-heavy users) can impair various cognitive processes such as memory and learning, attention control, fine motor control, and emotion processing~\cite{Kroon2021TheShort}. The use of cannabis in medical contexts (e.g., to treat chronic pain) has been examined in the
literature, including via observational studies~\cite{Boehnke2019PillsTo}; we focus here on 
cognitive aspects such as memory, fine motor control, decision-making, and creativity~\cite{Fridberg2010CognitiveMechanisms} that are related to programming~\cite{Siegmund2014UnderstandingUnderstanding}. 

Acute cannabis use impairs memory and learning as well inhibiting motor responses and reaction times~\cite{Kroon2021TheShort, Broyd2016AcuteAnd}. These cognitive processes are used while programming (e.g., during code comprehension tasks~\cite{Siegmund2014UnderstandingUnderstanding} or while typing code~\cite{Krueger2020NeurologicalDivide}). The impact of cannabis on other programming-related cognitive processes is less understood. Recent reviews report insufficient evidence of how cannabis impairs working memory or decision making~\cite{Kroon2021TheShort, Broyd2016AcuteAnd}, both essential to software~\cite{Crichton2021TheRole, Peitek2021ProgramComprehension, Siegmund2014UnderstandingUnderstanding}. Similarly, cannabis's impact on creativity is the subject of contradictory claims: For example, LaFrance \emph{et al.} found that cannabis users exhibit higher creativity due to increased openness to experiences~\cite{LaFrance2017InspiredBy} while Kowal \emph{et al.} found that cannabis impairs aspects of creativity at high dosages~\cite{Kowal2015CannabisAnd}. These conflicting claims preclude using a ``bottom up'' approach to infer the impact of cannabis on programming by considering its impact on relevant cognitive processes. We instead use a ``top down'' approach, studying the relationship between cannabis use and programming performance directly in a real-world context.

\textbf{Cannabis and Programming}. Endres \emph{et al.} surveyed 800 developers. Their results
provide a general context of the landscape of cannabis use in software engineering~\cite{Endres2022HashingIt}.  
They report that over one-third of their sample had used cannabis while programming and over one-sixth did so at least once per month. They also reported on cannabis use during work-related tasks. 
Key motivations were not medicinal but instead focused on ``perceived enhancement'' to software development
skills. Newman \emph{et al.} conducted interviews of 26 developers, identifying themes
related to soft skills, social stigma, 
and organizational policies~\cite{Newman2023FromOrganizations}. 
Critically, both of these prior studies rely on \emph{self-reported} impacts (e.g., cannabis-using
programmers say that cannabis enhances or hurts their programming). 

\textbf{Software Creativity}. Researchers have identified creativity as important to
multiple aspects of software engineering, from requirements elicitation~\cite{Nguyen2009AFramework}
to Agile development~\cite{Broderick2007EnhancingCreativity}. Creativity in software engineering is 
often associated with the novelty and effectiveness of the solutions~\cite{Mohanani2017Perceptions}  
and approaching problems from different angles~\cite{Groeneveld2021ExploringThe}.
Models and reviews 
have identified knowledge~\cite{Hedge2014HowTo}, communication~\cite{Groeneveld2021ExploringThe}, happiness~\cite{Graziotin2014HappySolve}, or personality~\cite{Amin2020TheImpact, Groeneveld2021ExploringThe} as factors relevant to creativity. 
Our study 
looks at divergent method-level implementation choice as one piece of creativity. 

\section{Experimental Setup and Design}

To understand the impact of cannabis intoxication on programming performance, we conduct a controlled observational study with 74 participants. Eligible participants were at least 21, had used cannabis in the last year, and had smoked or vaped cannabis before. We also required Python familiarity and programming experience comparable to that of a senior undergraduate. We first explain our study design in more detail. Then, in Section~\ref{subsec:stimuli}, we describe the surveys, programming tasks, and metrics used in our experiment.

\subsection{Study Design} 

\label{subsec:design}
We had three main design considerations: achieving sufficient statistical power for our pre-registered hypotheses (see Section~\ref{subsec:preRegistered}), balancing ecological validity with experimental control, and maximizing participant privacy and safety. 

\textbf{Overall Study Design.} 
To maintain statistical power, we use a within-subjects design. Participants completed a 20-minute questionnaire with demographics, cannabis usage history, programming history, and a four-minute training video introducing the study platform. Next, they attended two structurally-identical programming sessions: one cannabis-intoxicated and one sober.
The session order was counterbalanced: participants were randomly assigned the cannabis-first or sober-first condition (35/71 cannabis first vs. 36/71 sober first). This counterbalanced and within-subject design mitigates the impact of individual differences in programming ability, cannabis tolerance, and session ordering effects in our analysis.

\textbf{Study Session Structure.} All programming sessions lasted 1.5 hours and included two cognitive assessments, a series of short programming problems, and three ``interview-style'' coding questions.  We included cognitive assessments for data validation (see replication package), short programming problems for controlled observations of the impact of cannabis on programming, and ``interview-style'' coding questions to capture more complex coding algorithmic options. 
The short programming problems permit a controlled investigation of the impact of cannabis on specific aspects of programming while the ``interview-style'' questions enable a holistic and ecologically-valid analysis at the expense of statistical power and experimental control.
Section~\ref{subsec:stimuli} details all experimental tasks.

For the short programming problems, participants completed an online Qualtrics survey, a platform that permits randomization and timing collection via custom JavaScript. For the ``interview-style'' questions, participants wrote and executed code using a browser-based instance of VSCode (a popular programming text editor) via a Github Codespace configured to collect keystrokes, terminal/compiler interactions, and program file contents. Participants were also permitted to 
search (Google) for help with syntax errors. This design allowed our programming environment to have higher ecological validity, while remaining controlled enough to permit straight-forward statistical analysis. Sessions were conducted remotely (via Zoom) and participants were required to screen share.

\textbf{Cannabis Session Logistics.} Participants used cannabis 10--15 minutes before the start of the session, then uploaded pictures of the product and indicated the amount. Participants were instructed to consume cannabis via vaping or smoking, rather than by taking an edible to reduce variability: edibles can have very different effects than vaping or smoking~\cite{burt2021MechanismsOf}. 
We chose vaping based on its popularity in the the supplemental data from the Endres \emph{et al.} report~\cite{Endres2022HashingIt}. 

Participants who had used cannabis while programming before were asked to use the amount they would typically use when programming. If they had not, they were asked to use a mild to medium dosage, consistent with the amount they use when not programming. 
Allowing participants to choose the amount of cannabis to consume is different than the approach taken by many studies of cannabis use on behavior or cognition (cf.~\cite{Cuttler2021AcuteEffects}). 
However, allowing participants to self-select their usual cannabis dosage improves ecological validity. Informally, in this study we are not interested in learning the amount of impairment per milligram of cannabis, but instead in the amount of impairment an average cannabis-intoxicated programmer
faces while programming. We view the latter as significantly more actionable (e.g., to managers).

\textbf{Ethical Considerations.} The varying legality of cannabis requires special care in study design. Throughout the design and implementation of this study, we worked with our IRB (ethical review board) to ensure participant privacy and safety. Additional safety precautions were incorporated into our protocol including requiring that participants have used cannabis in the last year, and participating using a personally-owned computer (e.g., not a company-distributed laptop that may have tracking software) from a location where they would not have to travel for several hours. 

Ethical research practice also influenced our selection of a remote study design, rather than an in-person lab study. 
Design decisions such as directly administering specific amounts of 
cannabis at a central location or analyzing blood samples to assess intoxication 
were considered and rejected given our focus on ecological validity, as well as for reasons related to logistics and privacy. 

\subsection{Surveys and Stimuli: Content and Metrics}
\label{subsec:stimuli}

We describe our survey instruments, programming tasks, and metrics. 
All surveys and stimuli are 
in our replication package.\footnote{The replication package can be found through the OSF pre-registration (available here: \url{https://osf.io/g6fds}). A living repository for the project can be found on GitHub at \url{https://github.com/CelloCorgi/CannabisObservationalStudy}.}

\subsubsection{Demographics. }

We collected demographics such as gender, age, and employment status. We also asked programming experience, cannabis usage history, and prior experiences using cannabis while programming. Questions about programming experience and prior cannabis use while programming were adapted from a published survey~\cite{Endres2022HashingIt} to admit comparison with prior work. 
For general cannabis usage, we used the validated \textit{Daily Sessions, Frequency, Age of Onset, and Quantity of Cannabis Use Inventory} (DFAQ-CU)~\cite{Cuttler2017MeasuringCannabis}.

\if False 
\subsubsection{Cognitive Assessments}
\label{subsec:cognitive}

In each session, participants completed two common cognitive assessments: Stroop interference and a spatial ability test. The Stroop Effect is the longer reaction time humans typically have when presented with incongruent rather than congruent stimuli~\cite{Stroop1935StudiesOf}. The test we use involves colors: participants are shown a series of color words in various printed colors and are asked to respond to the printed color rather than the word meaning. In congruent stimuli, the word and the printed color are the same, while in incongruent stimuli they are different. We include the Stroop Effect in our study because it is a reliably-reproducible effect in psychology~\cite{Golden1978AManual} and thus it can be used to help validate the quality of our data. 

\emph{Spatial reasoning} is the capacity to understand, remember, and manipulate the orientation of objects in space, including both physical and abstract objects~\cite{Margulieux2019SpatialEncoding}. 
It is associated with mental animation, pattern recognition, and spatial perception. 
Spatial reasoning is in a family of tasks known to be impacted by cannabis 
intoxication~\cite{FIXME} and has been linked to STEM success in general
(e.g., ~\cite{Cheung2020SpatialMath, Stieff2007MRScience, Sorby2018DoesSpatial}) 
and to programming in particular~\cite{Cooper2015SpatialSkills,FIXME-SESMediatingSpatialSkillsPaper}.
The paper folding task is FIXME-MADDY~\cite{FIXME}. 

\textbf{Metrics and Scoring: } The Stroop effect is measured by subtracting each participant's average congruent response time from their average incongruent response time. Only correct response are considered with created average response times. The Paper Folding Test is multiple choice, and answers to individual questions are aggregated into a composite percentage score for analysis.

\fi

\subsubsection{Short Programming Problem Stimuli}
\label{subsec:shortProgramming}
\begin{figure}[t]
    \subfloat[Boolean problem (answer is False)]{
    \includegraphics[width=0.30\textwidth]{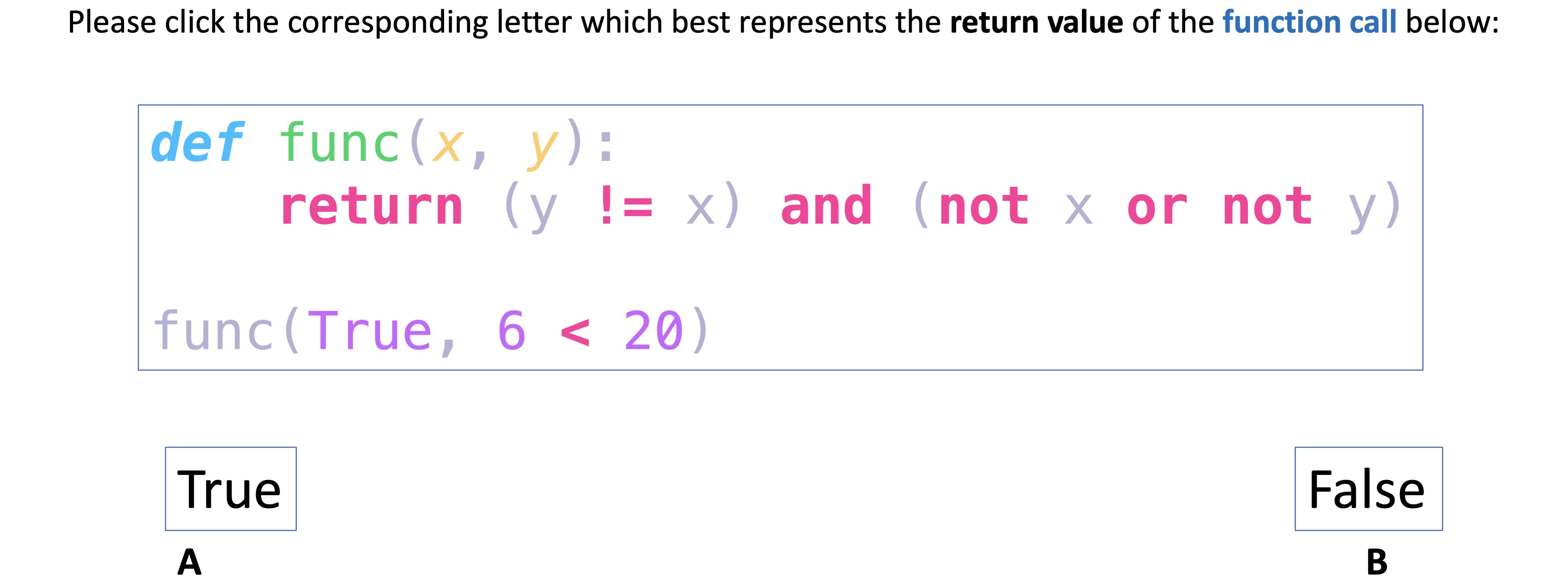}
    \label{fig:short-boolean}
    } \\
    \subfloat[Code-tracing problem (answer is 12)]
    {\includegraphics[width=0.30\textwidth]{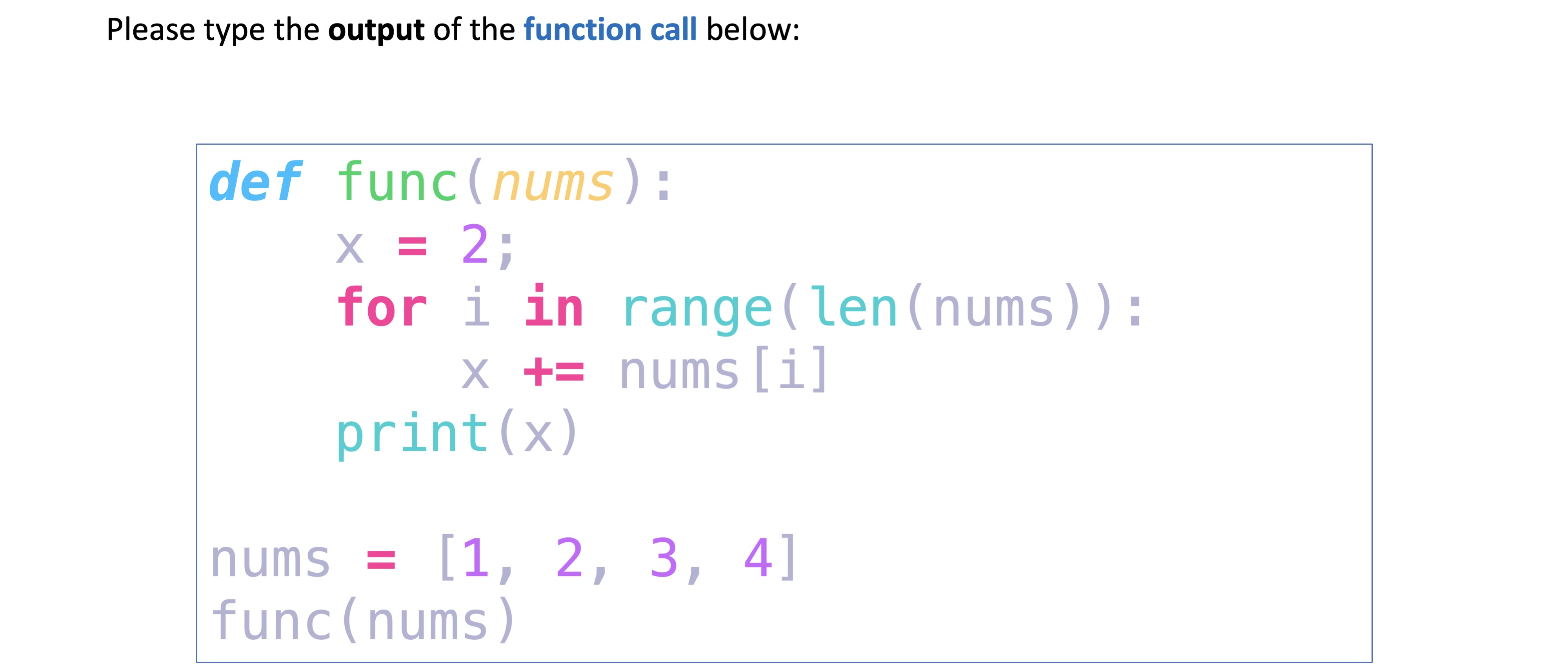}
    \label{fig:code-tracing}
    }\\
    \subfloat[Code writing problem]{
    \includegraphics[width=0.40\textwidth]{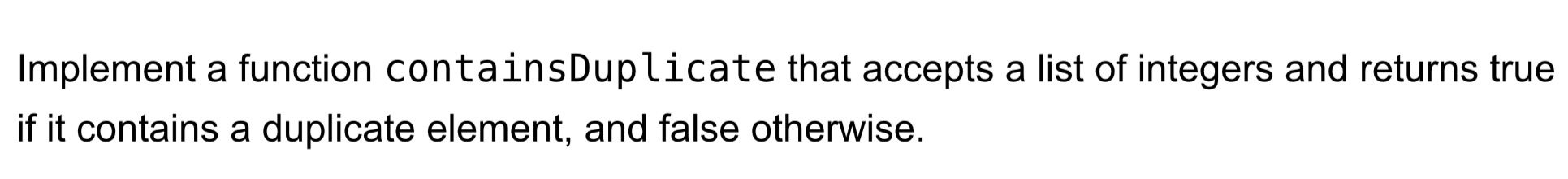}
    \label{fig:code-writing}
    } 
    
\caption{Example short programming stimuli, adapted from the program comprehension literature. }
\label{fig:short-stimuli}
\end{figure}

Each session included a series of short programming questions. We adapted 
stimuli from the programming comprehension literature, specifically those that use neuroimaging~\cite{Krueger2020NeurologicalDivide, Endres2021RelatingReading}. We chose to do this because such stimuli are explicitly designed to study targeted programming aspects in a controlled manner while still being completed quickly. We included three types of programming tasks: Boolean questions, code-tracing questions, and code-writing questions (see Figure~\ref{fig:short-stimuli} for examples). The Boolean and code-tracing stimuli were adapted from a study of novice programmer cognition~\cite{Endres2021RelatingReading}, while the code-writing stimuli were adapted from a study of code writing and prose writing~\cite{Krueger2020NeurologicalDivide}.

Consistent with their design and use in program comprehension studies, each stimulus was timed: participants had at most 30 seconds for each Boolean problem, 45 seconds for each code-tracing problem, and 90 seconds for each code-writing problem. In each session, each participant completed six problems from each sub-type that were randomly sampled from our corpus. 

\textbf{Metrics and Scoring.} Boolean problems were graded automatically. 
Code-tracing and code-writing problems were assessed manually by marking responses as either correct (full score), partially correct (half score), or incorrect (zero). Manual assessment of responses was done without knowing if the response was produced while high or sober. Our full rubric is available in our replication package. 
Graded responses were  aggregated into percent-correct values 
per sub-type per programming session for use in analysis. 

\subsubsection{``Programming Interview'' Style Stimuli}
\label{subsec:leetcode}

\begin{figure}[t]
    \includegraphics[width=0.40\textwidth]{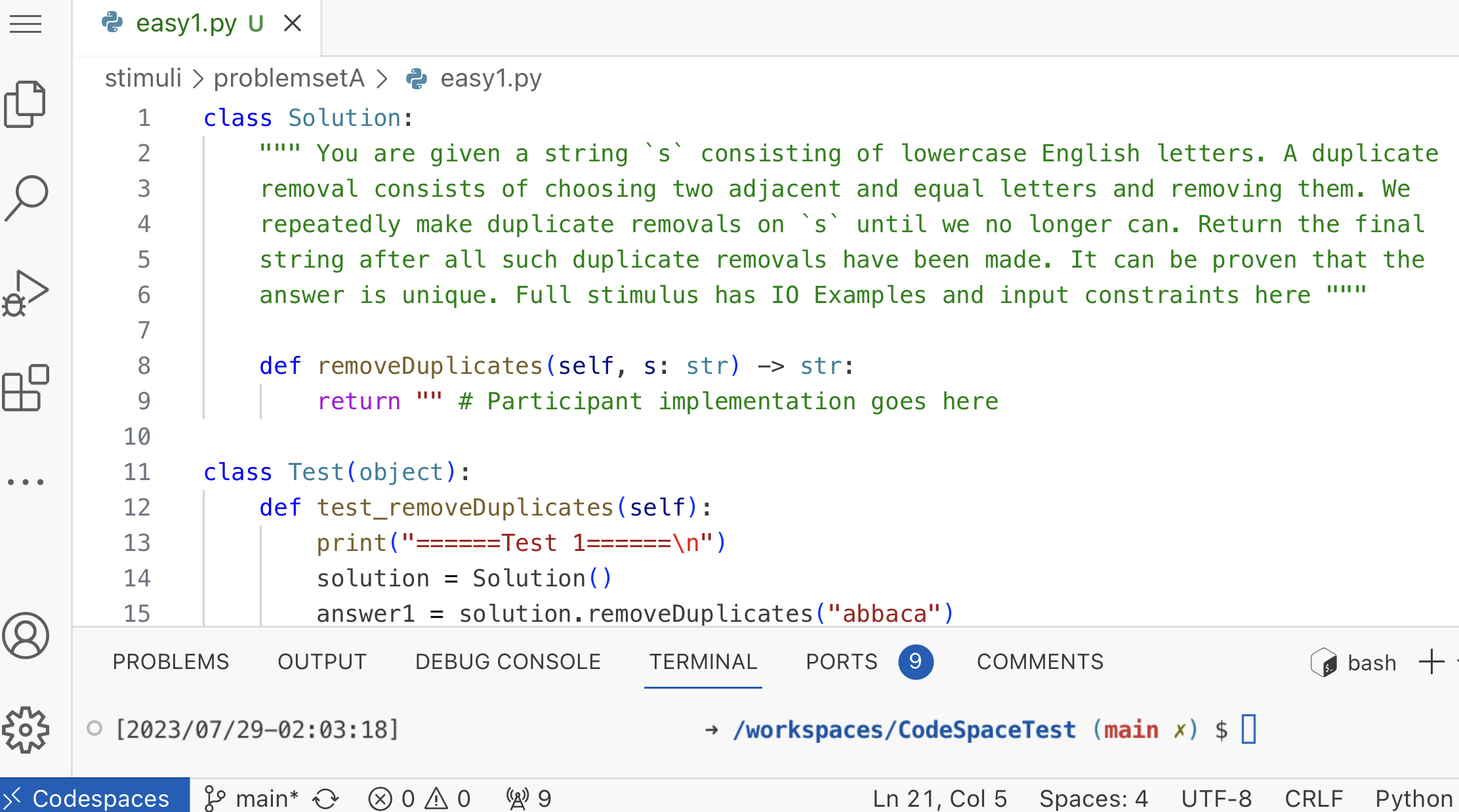}
    
\caption{Example ``interview-style'' programming stimulus, presented in the study platform (shortened for space). }
\label{fig:interview-stim}
\end{figure}

Each session included three ``Programming Interview'' problems. These were taken directly from LeetCode, a popular platform for practicing technical interview skills. Performance on these coding challenges has been found to reflect a software engineer's fundamental computer science knowledge, ability to find 
 efficient and scalable algorithms for unknown problems, and skill at testing or debugging a short piece of code~\cite{McDowell2015Craking}.
 Each problem consisted of a natural language specification, a function stub, and 2--3 basic tests. They admit implementation in a file where programmers can type, run, and edit code. We recorded the state of each program every 15 seconds, along with all keystrokes and terminal interactions. Figure~\ref{fig:interview-stim} shows an example of the study platform and an indicative stimulus.
 
To choose our stimuli, we first selected 20 potential problems labeled ``easy'' or ``medium'' on LeetCode. We avoided those marked ``hard'' due to the time constraints imposed by our study design. We categorized these problems into three groups by their primary data structure: a 1D-array, 2D-array, or a recursive data structure (list or tree).  
Through a series of pilot experiments, we selected six problems (two of each type) that were non-trivial but could be completed in the available time. In each session, participants completed one each of the two 1D-array, recursive, and 2D-array problems. A participant never saw the same problem twice. Problem order was randomized across sessions and study conditions to minimize between-problem variance and learning effect impacts. 
We partially controlled the difficulty of the problem pairs using the LeetCode problem difficulty and solve rate. For example, the two 1D-array and two recursive problems were ``easy'' on LeetCode, while the two 2D-array problems were marked as ``medium''. Participants had 15 minutes for each easy problem (1D-array, recursive) and 20 minutes for the medium problem (2D-array). 

\textbf{Automated Metrics and Scoring.} We assessed correctness via held-out test suites. As LeetCode does not publish its own hidden tests, we constructed our own held-out \emph{correctness tests}. 
The number of correctness test cases ranged from 24 to 34 per problem. 
All held-out test suites achieved full branch coverage on the published LeetCode solution. 
We analyzed the maximum correctness score across all saved file versions for a given solution. 
We use the best score rather than the final score because in cases where a participant ran out of time, the last score often is much lower because the participant was mid-edit. We also made 11 additional \textit{efficiency tests} per problem. 
These consisted of inputs of increasing size (including very large inputs), and were run separately from the correctness tests to ascertain the run-time efficiency of correct solutions. 




\textbf{Manual Annotation:} We qualitatively analyzed the 1D-array problem solutions to permit a nuanced analysis of design and code style factors that may differentiate high and sober programmers. We annotate algorithmic method choices (e.g., brute force, dynamic programming, etc.), and code style features (e.g., comments, helper functions, etc.). To determine 
algorithmic method categories, one author manually clustered the most up-voted Python solutions on 
LeetCode. This initial set was validated by another author. A third author manually assigned participant solutions into these clusters. Solutions about which the annotator was unsure were shown to the others and final categorizations were confirmed via consensus. 

\label{sec-2d-validation}
\textbf{Problem Difficulty}. We validated that the two alternate problems in each pairing had
similar difficulties for our population, regardless of LeetCode labels, as a potential source
of bias. We used a within-subjects $t$-test of participant scores. For the 1D-array and recursive pairings, we find no significant differences in difficulty ($p=0.82$, $0.66$ respectively). 
For the 2D-array problems, however, there is a significant difference in difficulty (68\% vs 34\% on average, $p<0.00001$). We consider this discrepancy when interpreting our results for the 2D-array problems in Sections~\ref{subsec:correctness} and~\ref{subsec:speed}.


\section{Participant Overview}

We overview our recruitment process and then describe the demographics, programming background, and cannabis usage history of our final $n=74$ participants to contextualize our results.

\subsection{Recruitment Process} 
\label{subsec:recruitment}
Participants were recruited via flyers posted around four American metropolitan areas (San Francisco Bay Area, Ann Arbor, Seattle, and New Haven). Each area is in a US state where recreational cannabis is legal (California, Michigan, Washington, and Connecticut). Within these areas, posters were primarily placed near the offices of technology companies and on university campuses. Prospective participants were directed to an online pre-screening form for eligibility requirements (Section~\ref{subsec:design}). This verification included eight programming questions to ensure sufficient Python performance. 
Of the 640 who completed the pre-screening, 247 obtained the perfect score required to participate. Participants were contacted in batches on a first-come-first-serve basis, with preference to those with more professional programming experience. In total, we sent 205/247 invitation emails before closing recruitment. 

Of the 205 invited, 85/205 finished the initial survey and scheduled programming sessions, a response rate of 42\%. 74 attended at least one session and 71 attended both, a study retention rate of 84\%. This overall rate aligns with previous software engineering studies with multiple sessions~\cite{Endres2021ToRead,Ford2022ATale}. Upon completion of the study, participants were compensated with an 80 USD gift-card to the company of their choice. All data collection occurred in 2023.

\subsection{Population Contextualization}

To better contextualize our results, we now describe the demographics, programming experiences, and cannabis use histories of our population. A summary of these numbers is available in Table~\ref{tab:demogrpahics}. 

\begin{table}[t]
\caption{Demographics\label{tab:demogrpahics}
and experience for $n=74$ participants. } 

\small
\begin{tabular}{@{}lr@{}}
\multicolumn{2}{l}{\textit{Gender}}                    \\ 
\midrule
Man                      & 53 (72\%)    \\
Woman                    & 15 (20\%)    \\
Non-binary               & 6 (8\%)      \\
\midrule
\multicolumn{2}{l}{\textit{Age and Programming Experience (Average, (Min--Max)) }}                    \\ \midrule 
                         
Age              & 24, (20--49)                 \\
Programming experience (years)  & 5,\textsuperscript{\ref{averagesMultChoiceExpl}} (1--30) \\ 
Has 1+ years of professional programming experience & 48 (65\%)\\
\midrule
\multicolumn{2}{l}{\textit{Computing-related employment status (could select multiple)}}                    \\ \midrule
Currently Employed at a CS-related job       & 28 (38\%)    \\
Undergraduate Student in CS related field   & 37 (50\%) \\
Graduate Student in CS-related field & 12 (16\%)  \\
Unemployed or N/A         & 3 (4\%)     \\ 
\bottomrule
\end{tabular}
\end{table}

Participants ranged in age from 20 to 49 (average 24), with 72\% men, 15\% women, and
8\% non-binary. 
Our population had a mix of students and full time professional developers. About half (37/74) were undergraduates in a computing field. The remainder were either graduate students (16\%, 12/74) or 
professional programmers (38\%, 28/74). A few reported both student status and current programming employment. This split between students and professionals is consistent with the locations we put up recruitment posters.

Participants reported a range of programming experiences. While all had at least one year of programming experience (a requirement to participate), eight participants had over 10 years of experience, and the median was 5.\footnote{\label{averagesMultChoiceExpl} Estimate: Years of experiences was reported in ranges (e.g., 1-2, 3-5, 6-10, etc.)} As for professional programming experience, participants ranged between none and over 20 years. The majority of participants in our sample had professional programming experience (94\%), with the median participant reporting between 1 and 5 years: 65\% had at least one year of professional experience. 
Of those who had at least one year, the most common job titles reported were ``software engineer'' or ``software developer''.

\textbf{Cannabis Usage History.} All eligible participants had used cannabis in the last year. Specific usage levels varied substantially, ranging from once a year to more than once every day. The median participant reported using twice a week. In addition, the majority of our participants had experience programming while high (66\%, 49/74). The frequency of this use ranged from under once a year to five or more days a week. The wide array of cannabis usage histories, both with programming and without, allows us to systematically investigate if the magnitude of prior cannabis use mediates the impacts of cannabis intoxication on programming performance. 


\section{Research Questions and Analysis}

We structure our analysis in two parts: a primary hypothesis-driven analysis and a secondary exploratory analysis. The primary hypotheses and analysis plan was preregistered to mitigate biases and increase confidence in our results. However, as investigating the impact of cannabis on programming is relatively novel, we perform additional exploratory analysis to glean insights for further study. In the rest of this section, we present our research questions for both analysis parts, and  outline our statistical methods.

\subsection{Primary Analysis: Pre-registered Questions}
\label{subsec:preRegistered}
Following more recent best practices in both software engineering (e.g., the ``Registered Reports'' track of Mining Software Repositories~\cite{msr2023}) 
and psychology and the social sciences (e.g.,~\cite{Simmons2020PreregistrationWhy}), we
\emph{pre-registered} our hypotheses before conducting our analysis. 

In pre-registration, the ``research rationale, hypotheses, design and analytic strategy'' are submitted before beginning the study~\cite{Gonzales2015ThePromise}. 
As a result, biases associated with researchers 
choosing which results to present after the fact may be mitigated:
``pre-registration can prevent or suppress HARKing, p-hacking, and cherry picking since hypotheses and analytical methods have already been declared before experiments are performed''~\cite{Yamada2018HowTo}.
Similarly, pre-registration may mean that 
``researchers will not be motivated to engage in practices that
increase the likelihood of making a type I error''~\cite{Gonzales2015ThePromise}.

For this study, we pre-registered four research questions and associated hypotheses with the Open Science Framework (OSF), along with our data collection strategy and statistical analysis methods:\footnote{OSF pre-registration is available here: \url{https://osf.io/g6fds}. We note that in the pre-registration, \textit{RQ3} mentions \textit{creativity} instead of \textit{program method divergence}. As suggested by our reviewers, we change the name in this paper to better match our methods. }

\begin{itemize}[leftmargin=15pt,topsep=2pt]
    \item \textbf{RQ1---Program Correctness:} How does cannabis intoxication while programming impact program correctness?
    \begin{itemize}
        \item \textit{Hypothesis:} Programs will be less correct when written by cannabis-intoxicated programmers.
    \end{itemize}
    \item \textbf{RQ2---Program Speed:} How does cannabis intoxication while programming impact programming speed?
    \begin{itemize}
        \item \textit{Hypothesis:} Cannabis-intoxicated programmers will take longer to write programs. 
    \end{itemize}
    \item \textbf{RQ3---Program Method Divergence:} Does cannabis use influence programmer algorithmic method choice?
    \begin{itemize}
        \item \textit{Hypothesis 1: } Correct programs by high participants will run slower than those by sober participants (i.e. are less efficient or have higher algorithmic complexity).\footnote{In our pre-registered hypotheses, this was listed under RQ2. We present it under RQ3 for thematic and narrative clarity.}
        \item \textit{Hypothesis 2:} Solutions to free-form programming problems by cannabis-intoxicated programmers will exhibit greater method choice divergence and diversity.
    \end{itemize}
    \item \textbf{RQ4---Cannabis Use History:} Does cannabis usage history mediate intoxication's
    on programming outcomes? 

    \begin{itemize}
        \item \textit{Hypothesis:} The impact of cannabis use while programming will be lessened for heavy vs. moderate users.
    \end{itemize}
    
\end{itemize}

\subsection{Exploratory Analysis}
\label{subsec:exploratory}

We also consider two exploratory research questions:

\begin{itemize}[leftmargin=15pt,topsep=2pt]
    \item \textbf{E-RQ1---Code Style:} Does cannabis intoxication impact stylistic code properties (e.g., code comments, etc.)?
    \item \textbf{E-RQ2---Self Perception:} Are programmers able to accurately assess how cannabis impacts  programming performance?

\end{itemize}

\subsection{Statistical Methods}
\label{subsec:stats}

Our analysis was primarily conducted in a Python Jupyter Notebook 
using \texttt{Pandas}~\cite{pandas}. Some analyses, especially those informed by data visualization, were done using Excel. For statistical tests, we primarily used the \texttt{SciPy}~\cite{2020SciPy-NMeth} and \texttt{Statsmodels}~\cite{seabold2010statsmodels} libraries. 

\textbf{Significance.} We consider results significant if $p<0.05$. When testing for a significant difference between sober and cannabis-intoxicated programming using continuous variables (e.g., percent correctness scores or response time such as in RQ1 and RQ2), we use a paired samples $t$-test unless otherwise noted. 
Assumptions of normality are confirmed through inspection of histograms.
While we primarily use paired tests (as is appropriate with our within-subjects design), in some
cases (e.g., missing data, etc.) we use a non-paired test and note the specific test used in the text. 



For categorical values (e.g., program method choice in RQ3), we use a $\chi^2$-test. For the difference between two binary variables (such as if a solution has comments or not in E-RQ1), we use the $n$-1 $\chi^2$-test (i.e., the proportions $z$-test)~\cite{campbell2007chi}.
We treat the responses to Likert questions (e.g., self-perception of cannabis impact in E-RQ2) as continuous variables.\footnote{Although they are ordered categorical variables and normality cannot be assumed, with large samples, parametric tests are sufficiently robust for analysis~\cite{likertStatistics}.} The Student's $t$-test is thus appropriate.

\textbf{Multiple Comparisons.} We investigate multiple research questions and conduct multiple statistical tests per research question. To avoid fishing or $p$-hacking, we pre-registered our primary hypotheses and analysis plan and report results for each. 
Within each research question, we correct for multiple comparisons for all tests used to accept or reject the null hypothesis. We use Benjamini-Hochberg Correction, with a false discovery threshold of $q=0.05$: unless stated otherwise, all significant results pass this threshold.

\textbf{Effect Size.} We use Cohen's $d$ (with pooled standard deviation) to assess the size of differences tested by $t$-tests. We consider $d >0.2$ a small effect, and values above $0.5$ a medium effect. For correlations, we use Pearson's $r$, with $0.1 < r \le 0.3$ a weak correlation, $0.3 < r \le 0.5$ a moderate correlation and $0.5 < r$ a strong correlation. 

\section{Results}

We present the results of our pre-registered (RQ1-RQ4 Section~\ref{subsec:preRegistered}) and exploratory (E-RQ1, E-RQ2, Section~\ref{subsec:exploratory}) questions. 
Section~\ref{subsec:stimuli} details experimental tasks, including the metrics used. 
A discussion of the statistical our statistical methods is in Section~\ref{subsec:stats}.

\subsection{RQ1 --- Impacts on Program Correctness}
\label{subsec:correctness}

We first investigate how programming while intoxicated impacts program correctness. We do this by using paired $t$-tests to compare sober vs. cannabis session percent correctness for each short program comprehension task type and ``interview style'' coding question. This results in six total significance tests. 

\begin{table*}[]
\small
\caption{The impact of cannabis intoxication on programming correctness (RQ1) and speed (RQ2). 
\rm{All significance tests are paired $t$-tests, while effect size calculations use Cohen's $d$. Cells that are bold and highlighted in green are those differences that are significant with Benjamini-Hochberg correction ($p$< 0.05, $q$=0.05). Cells in italics and highlighted in yellow indicate a trend that did not reach significance. Notice that all differences, even those that did not reach significant, indicate decreased performance while high: 
\label{tab:RQ1AndRQ2TopLevel}
cannabis-intoxicated programmers take more time to write more incorrect programs.} } 
\begin{tabular}{@{}lrrrrrr|rrrrrr@{}}

\toprule
                & \textbf{Sober} & \textbf{High} & \textbf{Diff}                   & \textbf{p}       & \textbf{BH-p}                          & \textbf{d} & \textbf{Sober} & \textbf{High} & \textbf{Diff}                & \textbf{p}                             & \textbf{BH-p}                          & \textbf{d} \\ \midrule
\multicolumn{7}{l|}{\textit{RQ1:  Short Programming Problems, Correctness Scores}}                                                                          & \multicolumn{6}{l}{\textit{RQ2: Average Stimulus Time (in seconds)}}                                                                                         \\ \midrule
Boolean         & 81.5\%         & 81.0\%        & -0.5\%                          & 0.846            & 0.846                                  & 0.03       & 14.2           & 14.7          & +0.5                         & 0.310                                  & 0.465                                  & 0.10       \\
Code-tracing    & 62.3\%         & 52.1\%        & \cellcolor[HTML]{FF7F7C}-10.2\% & \textless{}0.001 & \cellcolor[HTML]{A7FF8B}\textbf{0.003} & 0.42       & 31.7           & 32.1          & +0.4                         & 0.656                                  & 0.656                                  & 0.06       \\
Code-writing    & 56.9\%         & 46.4\%        & \cellcolor[HTML]{FF7F7C}-10.6\% & \textless{}0.001 & \cellcolor[HTML]{A7FF8B}\textbf{0.003} & 0.44       & 67.4           & 70.4          & \cellcolor[HTML]{FFCCC9}+3.0 & \cellcolor[HTML]{FFFFC7}\textit{0.065} & \cellcolor[HTML]{FFFFC7}\textit{0.130} & 0.23       \\ \midrule
\multicolumn{7}{l|}{\textit{RQ1: ``Programming Interview'' Problems, Correctness Scores}}                                                                   & \multicolumn{6}{l}{\textit{RQ2: Average Overall Time (in min)}}                                                                                              \\ \midrule
Problem 1: Strings and 1D arrays     & 65.9\%         & 56.4\%        & \cellcolor[HTML]{FF7F7C}-9.5\%  & 0.033            & \cellcolor[HTML]{A7FF8B}\textbf{0.049} & 0.28       & 9.7            & 11.1          & \cellcolor[HTML]{FF7F7C}+1.4 & 0.012                                  & \cellcolor[HTML]{A7FF8B}\textbf{0.039} & 0.32       \\
Problem 2: Recursive Lists and Trees& 48.5\%         & 34.5\%        & \cellcolor[HTML]{FF7F7C}-14.0\% & 0.012            & \cellcolor[HTML]{A7FF8B}\textbf{0.024} & 0.35       & 11.8           & 13.0          & \cellcolor[HTML]{FF7F7C}+1.2 & 0.013                                  & \cellcolor[HTML]{A7FF8B}\textbf{0.039} & 0.33       \\
Problem 3: 2D-arrays          & 53.9\%         & 48.4\%        & \cellcolor[HTML]{FFCCC9}-5.5\%  & 0.383            & 0.460                                  & 0.15       & 16.2           & 16.8          & \cellcolor[HTML]{FFCCC9}+0.6 & 0.500                                  & 0.656                                  & 0.11       \\ \bottomrule
\end{tabular}

\end{table*}

At a high level, we find strong evidence supporting our hypothesis that cannabis-intoxicated written programs are less correct. For all correctness score comparisons, participants' cannabis correctness scores were \textit{lower}, on average, than sober scores. For four out of the six comparisons, this difference was statistically significant ($0.0005<p<0.05$) with small to small-medium effect ($0.28\leq d \leq0.44$). Table~\ref{tab:RQ1AndRQ2TopLevel} summarizes our top-level results. 

\begin{figure}[t]
    \begin{subfigure}{0.35\textwidth}
        \label{fig:codeSober}
        \caption{Code produced by participant when sober}
        \begin{lstlisting}
  def is_sorted(integers):
    for i in range(len(integers)-1):
      if integers[i] > integers[i+1]:
        return False
    return True
        \end{lstlisting}

    \end{subfigure}
    
    \begin{subfigure}{0.35\textwidth}
        \label{fig:codeIntoxicated}
        \caption{Code from same participant when intoxicated}
            \begin{lstlisting}
  def is_sorted(input_list):
    return helper(None, input_list)
    
  def helper(min_val, input_list):
    if len(input_list) == 0: return True
    if min_val > input_list[0]: return False
    return helper(input_list[0], input_list[1:])
        \end{lstlisting}        
    \end{subfigure}

    \caption{Indicative example comparing code produced while high vs. sober by the same participant for the same problem.
    The intoxicated code is more complicated and contains a bug.}
    \label{fig:shortProgComparison} 
\end{figure}

\textbf{Short Programming Problems. } Participants completed several Boolean logic problems, code-tracing problems, and code-writing problems during each session. 
For the Boolean task, we do not find evidence of cannabis intoxication impairing performance ($p=0.85$).  
However, for both the code-tracing task and the code-writing task, we find that cannabis intoxication has a significant negative impact on performance ($p=0.0009$, $p=0.0005$ respectively). For both of these tasks, the negative impact was of a small-medium effect ($d=0.42$, $d=0.44$). To put this in perspective, correctness scores were, on average, 10\% lower when intoxicated (52\% vs. 62\% for code-tracing and 46\% vs. 57\% for code-writing). Cannabis impairment at ecologically-valid levels can have a negative impact on two fundamental software development tasks: code reading via tracing and code writing. Additionally, the presence of an effect for the two more cognitively-demanding tasks, but not for the simpler Boolean task, may indicate that cannabis intoxication impacts scale with task complexity; if so, this trend could relate to known effects of cannabis on working memory~\cite{Kroon2021TheShort}. 

To understand the factors driving this decreased performance, we perform an informal qualitative analysis of the types of errors that cannabis-intoxicated programmers are likely to make on the code-tracing and code-writing tasks. We observe that high programmers often complicate their solutions and add extraneous conditionals while still missing edge cases. 
Figure~\ref{fig:shortProgComparison} shows an indicative example: code produced by the same participant for the same problem (on different days) while high and sober. The code produced while intoxicated features
a more complicated structure (recursion via a helper function) as well as more opportunities for simple mistakes (e.g., three indexing operations vs. two, two conditional branches vs. one, etc.). The intoxicated solution contains one such error, comparing \texttt{None} with a number, which raises a \texttt{TypeError}.



\textbf{``Interview style'' Problems.} While the short problems highlight the impact of cannabis on specific programming aspects, for a more holistic understanding we consider the three longer ``interview style'' problems.
Participants never completed the same interview problem twice: problems were paired and counterbalanced across sessions by algorithm type and difficulty to permit within-subjects analysis. The first problem always involved 1D-arrays, the second recursive trees or lists, and the third 2D-arrays. 

We find that cannabis significantly impaired correctness for the 1D-array problems and recursive problems. For the 1D-array problems, participants passed 10\% fewer correctness tests in the cannabis-intoxicated session than in the sober session (56\% vs 65\%, $p = 0.033, d = 0.28$). For the recursive problems, participants passed 14\% fewer correctness tests in the cannabis-intoxicated session than in the sober session (34\% vs. 48\%, $p = 0.012, d = 0.35$). 

For the 2D-array problems, while participants passed 6\% fewer correctness tests in the cannabis session, this trend did not rise to the level of significance (48\% vs 54\%, $p = 0.383, d = 0.15$). We note, however, that unlike the 1D-array and recursive problems, the two 2D-array problems that we chose were not of equivalent difficulty for our population (Section~\ref{sec-2d-validation}). This confound may 
explain the lack of an observed significant difference on this problem pair.

In our pre-registered analysis plan, we stated that ``The null hypothesis will be rejected if high programmers have lower scores on the majority of interview-style\footnote{The original references LeetCode explicitly. We use interview-style in this paper.} problems when high.'' As we found a significant difference in 2/3 interview-style problems, we reject the null hypothesis and conclude that cannabis intoxication has a significant negative effect on program correctness. Specifically, this result shows that the impairment we observed on the controlled short programming tasks persists when implementing more complicated functions. We discuss implications in Section~\ref{sec:discussion}.

\begin{tcolorbox}[colback=palegreen, colframe=black, colbacktitle=white, coltitle=black, boxsep=0pt,top=6pt,bottom=6pt,left=6pt,right=6pt, beforeafter skip=6pt]
    \noindent We find support for our hypothesis that 
    \textbf{cannabis use decreases program correctness} with a small-medium effect ($0.0005 < p < 0.05$, $0.28\leq d \leq0.44$, 10--14\% fewer passed tests). Cannabis impairs \textit{writing} and \textit{tracing} through programs. 

\end{tcolorbox}

\subsection{RQ2 --- Impacts on Programming Speed}

\label{subsec:speed}

We next investigate the impact of cannabis on programming speed. We hypothesized that cannabis-intoxicated people will program more slowly. 
For the short programming problems, we use a paired $t$-test to compare the average stimulus completion time per task type (Boolean, code-tracing, code-writing) per condition (sober or intoxicated). 
For the ``interview-style'' questions, we compare each problem type's total completion time. Our results are in Table~\ref{tab:RQ1AndRQ2TopLevel}.

\textbf{Short Programming Problems.} We find no significant evidence that cannabis intoxication impacts programming time for simpler programming tasks. For the code-writing task (the most complicated of the three), we observe a trend toward significance with task completion taking longer when intoxicated with small effect ($p=0.065$, $d=0.23$). However, this difference neither  reaches significance nor passes our multiple comparison threshold.

\textbf{``Interview Style'' Problems.} In contrast, we do find a significant programming speed difference for 2/3 of the more complex ``interview style'' problems. For the 1D-array problems, high participants spent an average of 84 more seconds than sober participants (11.1 minutes vs. 9.7 minutes, $p = 0.01, d = 0.32$). For the recursive problems, high participants spent an average of 83 more seconds than sober participants (13.0 minutes vs. 11.8 minutes, $p = 0.01$, $d = 0.33$). We do not observe a significant difference between completion times for the 2D-array problems. However, as with correctness (Section~\ref{subsec:correctness}), this lack of result may be attributable to uneven difficulty matching for this problem pair (Section~\ref{sec-2d-validation}).

\textbf{Why Are Programmers Slower?} We investigate the factors driving the difference for the ``interview style'' questions: are intoxicated programmers slower because they physically type slower, because they make more typing corrections, or because they have less ``active typing time'' (time spent not actively programming, but rather thinking or searching online for help)? 

\begin{figure}[t]

    \includegraphics[width=0.46\textwidth]{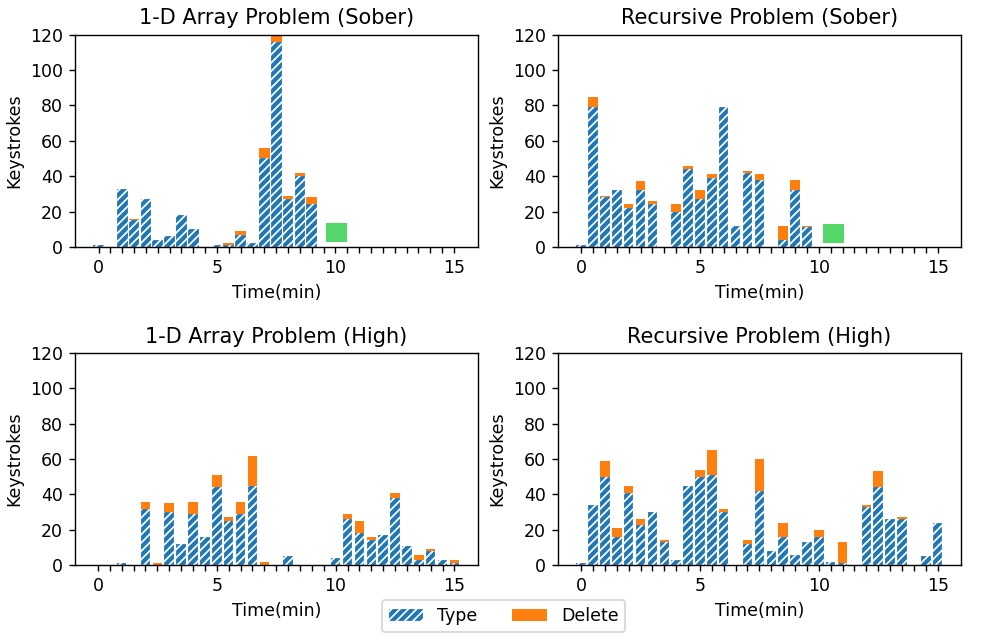}
    \label{fig:patterns}
\caption{Histograms with typed characters (dark blue with lines) and deletions (orange) over time for
\label{fig:sessionTyping}
the same participant while sober and high. 
\rm{ 
The high condition features longer pauses and more deletions. The participant also had a higher maximum typing speed sober (120 keystrokes in 30 seconds vs. only 65 while high). This participant finished both problems early sober (noted by the green box) but ran out of time when high.}
}

\end{figure}

We find that all three factors contribute! 
We compared participants' overall typing speed in both sessions using a paired $t$-test.
Sober participants typed 6 more characters per minute than cannabis-intoxicated participants on average (84 chars/min vs. 90 chars/min, $p = 0.0004, d = 0.32$). We excluded the time when participants were not actively typing, defined as any period between two keystrokes longer than 8s, 
from the total time used to calculate typing speed. 
For corrections, cannabis-intoxicated participants delete more of their keystrokes than sober participants (20.9\% vs. 18.5\%, $p = 0.00003, d = 0.35$). This result, along with the slower typing speed, aligns with work on general cannabis impairment, which finds a negative impact on fine motor control~\cite{Kroon2021TheShort}. 
Finally, we find that cannabis-intoxicated participants spend more of their total time not actively typing code (64.9\% vs. 60.6\%, $p = 0.003, d = 0.36$). We visualize these typing-related differences between the sober and cannabis sessions for a single indicative participant in Figure~\ref{fig:sessionTyping}.

\if False
Pre-registered plan
For this one, we will run three tests:
1) Typing speed. We will compare the average characters typed per minute between high participants and sober participants during the leet code problems using a t-test. We expect it to be lower for high participants.
2) We will look only at those participants who got a correct leet code or short-answer survey coding solution (For leet-code, they must have gotten at least 75\% of the held-out tests). We will then, per problem, use a t-test to compare how long it took to get to that solution.
3) We will check how long the held-out tests took to run for those participants that got all problems correct. We expect the held out tests to take longer to run for high participants.
\fi

\begin{tcolorbox}[colback=palegreen, colframe=black, colbacktitle=white, coltitle=black, boxsep=0pt,top=6pt,bottom=6pt,left=6pt,right=6pt, beforeafter skip=6pt]
    \noindent For ``interview-style'' tasks,
    \textbf{cannabis use impairs programming speed} ($p < 0.04$, $d = 0.3$, 10--14\% slower). 
    This decrease in speed is associated with typing slower, deleting more characters, and more time spent not typing. 
\end{tcolorbox}

\subsection{RQ3 --- Method Choice and Divergence}


We now investigate if high and sober programmers \textit{choose to solve the same programming problem in different ways}. We consider both the \textit{efficiency} of solutions and also the \textit{algorithmic method} implemented. 
We have two hypotheses: first, correct programs by high participants will be less efficient than those of sober participants. Second, we hypothesize high participants will show more divergence choices 
in algorithmic or methodological approaches
(one potential aspect of ``creativity''), compared to sober participants.

We focus on the ``interview-style'' problems, as those tasks are complex enough for a meaningful algorithmic analysis. For RQ3 we analyze all six problems separately (i.e., two 1D-array, two recursion, and two 2D-array problems). This is done because the running times of difficulty-paired problems may vary significantly. For example, for the 1D-array problem type, one instance features a 1D-array as a Python list while the other uses a Python string. While conceptually similar, Python treats these very differently from an efficiency standpoint (e.g., list instances are mutable while strings are not). The statistical comparisons in this research question thus use non-paired tests unless otherwise noted.

\textbf{Program Efficiency.} We measure the efficiency of correct solutions on very large program inputs (Section~\ref{subsec:leetcode}, average of three trials). All program running times, 
were generated using the same multi-core Linux server and were run sequentially. 
Despite generous timeouts for the efficiency tests, some particularly-inefficient correct solutions failed to terminate for our biggest inputs. For these problems, we assign them our maximum timeout of 60 seconds. 

We compare the efficiency test runtimes for correct solutions written while high to those written while sober for each of the six ``interview-style problems''. For 5/6 problems, the differences are not significant. For one of the two 1D-array problems, the difference is significant before multiple comparison correction ($p=0.03$, $d=1.03$). For this problem, the solutions by sober participants are actually \textit{less efficient} than those by high participants (5.0 seconds vs. 21.7 seconds). While intriguing, this result does not survive correction for multiple comparisons (corrected $p=0.18$). Additionally, few high and sober participants correctly implemented the problem (9 and 12 respectively), leading to low statistical power. We cannot reject the null hypothesis regarding efficiency. 

\textbf{Solution Divergence.} We manually annotate the solutions to both 1D-array problems to obtain a more nuanced understanding of method choice differences between high and sober participants. 
We note that, unlike for our efficiency analysis, we manually annotate the solutions for all participants, even those who did not arrive at the correct solution. This is so 
because there is anecdotal evidence that cannabis use might improve programming \textit{creativity}~\cite{Endres2022HashingIt, Newman2023FromOrganizations}. If this is the case, even if solutions produced by cannabis-intoxicated programmers contain more bugs, other benefits may offset this cost; informally, a developer might generate a solution while using cannabis and then come back the next day to fix any errors. 

To the best of our knowledge, while creativity is an important aspect of the software development process, a robust metric for it remains an open problem~\cite{Groeneveld2021ExploringThe}.  One approach used by prior work is to measure \emph{divergence} in computational patterns~\cite{Bennett2013ComputingCreativity}. Divergence tests have long been the basis of common creativity measurements~\cite{Runco1988Problem}. We adapt this use of solution divergence to method choice. 

The possible method choices were specific to the first or second problem instance, and included categories such as brute force using a loop, brute force using recursion, or a stack data structure. For both problem instances, a couple of submissions did not fall into any category and were labeled other, or were categorized as completely incorrect. Section~\ref{subsec:leetcode} overviews our method annotation process in more detail. We applied a $\chi^2$ test with the groups as sober and intoxicated and the categories as the different
methods for each problem. For the first problem instance, $\chi^2 = 1.68, p = 0.89$. For
the second problem instance, $\chi^2 = 8.44, p = 0.077$. We find no evidence that method choice differed significantly between high and sober participants. As a result, it does not make sense to investigate if methods chosen by high participants were more diverse because there was no significant difference between the two distributions. Overall, we find no evidence to support our hypothesis that high programmers generate more diverse or more creative programming solutions. We discuss implications in Section~\ref{sec:discussion}.

\begin{tcolorbox}[colback=paleblue, colframe=black, colbacktitle=white, coltitle=black, boxsep=0pt,top=6pt,bottom=6pt,left=6pt,right=6pt, beforeafter skip=6pt]
    \noindent We found \textbf{no statistically-significant evidence that cannabis intoxication impacts solution efficiency or implementation divergence} ($p \ge 0.08$). 
    We do not reject the null hypothesis that programmers using cannabis exhibit the same divergence of method choice as sober counterparts. 

\end{tcolorbox} 

\subsection{RQ 4 --- Influence of Cannabis Use History} For our last pre-registered hypothesis, we investigate if cannabis use history mediates the negative impact of cannabis intoxication on program correctness. 
We hypothesize the impact of cannabis use while programming will be lessened for those that are heavy cannabis users vs. those that are moderate users. 

We divide participants into two groups for analysis: heavy users and light users, classified by if their aggregated $z-$transformed scores on the DFAQ-CU use frequency questions are positive or negative~\cite{cuttler2017measuring}.\footnote{In the pre-registered hypotheses, we said we would use three groups: light cannabis users (at most 3--4 times per month), moderate cannabis users (1--2 times per week), and heavy cannabis users ($2+$ times per week). We use an aggregated score instead after a closer inspection of the DFAQ-CU assessment's scoring instructions~\cite{Cuttler2017MeasuringCannabis}.} 38 participants are classified as light users while 33 are classified as heavy users (roughly, those participants who use cannabis more than two times a week). We then calculate the per-participant difference between high and sober sessions for all correctness scores for which we observed significant general impairment: code-tracing problems, code-writing problems, the 1D-array problem type, and the recursive-data structure problem type. We use an independent $t$-test to compare the performance differences of light and heavy users on each test to see if one group experiences significantly more cannabis-related impairment than the other. 

We find no significant differences in cannabis-related impairment between heavy and light users ($0.35 \leq p \leq 0.88$). To confirm this null result, we additionally compute pairwise correlations between inter-session performance correctness differences and two cannabis-related features: self-reported current intoxication level and life-time cannabis usage amount. We find no significant correlations with these comparisons ($ -0.26 < r <0.13$ for all correlations).

\begin{tcolorbox}[colback=paleblue, colframe=black, colbacktitle=white, coltitle=black, boxsep=0pt,top=6pt,bottom=6pt,left=6pt,right=6pt, beforeafter skip=6pt]
    \noindent We find \textbf{no significant evidence that cannabis impacts heavy users less than others} ($p \ge 0.35$). We do not reject the null hypothesis, and instead conclude that cannabis impairs all programmers equally, regardless of cannabis use history.

\end{tcolorbox}

\subsection{E-RQ1 --- Impacts on Code Style}

We explore the impact of cannabis use on code style. While annotating participant method choices for the two 1D-array problems, we also annotated responses for various stylistics features. In particular, we marked if a participant added comments, print statements, helper functions, or additional test cases to their implementation. We also counted the number and maximum nesting depth of loops and conditionals to get a measure of branch and loop complexity of the code. 
We compare proportions for stylistic features between high and sober participants using the $n$-1 $\chi^2$-test. This analysis 
determines if the correctness impairment of cannabis extends to stylistic properties which, while non-functional, facilitate software readability and maintainability~\cite{Fakhoury2018TheEffect,Fry2012AHuman}. 

We find no significant style differences  between programs written while high vs. sober ($0.20 \leq p \leq 0.85$). While exploratory, this may mitigate concerns about cannabis substantially compromising code clarity and ease of understanding, which are critical for successful collaboration and future program modifications.

\if False

 Here are the results for comments (aggregated across both problems) groups: high a
nd not high, categories: Has At Least One Comment and No Comments
    HasAtLeastOneComment NoComments
High 
Not High
$\chi^2 = 0.0366, p = 0.8484$
 Manasvi question .... one trend - MaxNestedLoops for both problems have relatively low p values (0.2467, 0.1973) and the submissions who were high had higher average across both problems compared to submissions who were sober (1.514 compared to 1.361) and (1.667 compared to 1.514)--> to Wes' point this has marginal maintainability impacts?
\fi

\begin{tcolorbox}[colback=paleblue, colframe=black, colbacktitle=white, coltitle=black, boxsep=0pt,top=6pt,bottom=6pt,left=6pt,right=6pt, beforeafter skip=6pt]
    \noindent We find \textbf{no significant evidence that cannabis impacts programming style} (e.g., comments, helpers, etc.) ($p \ge 0.20$).  

\end{tcolorbox}

\subsection{E-RQ2 --- Self Perception of Impact}

\begin{figure}[h]
\centering
\includegraphics[width=0.35\textwidth]{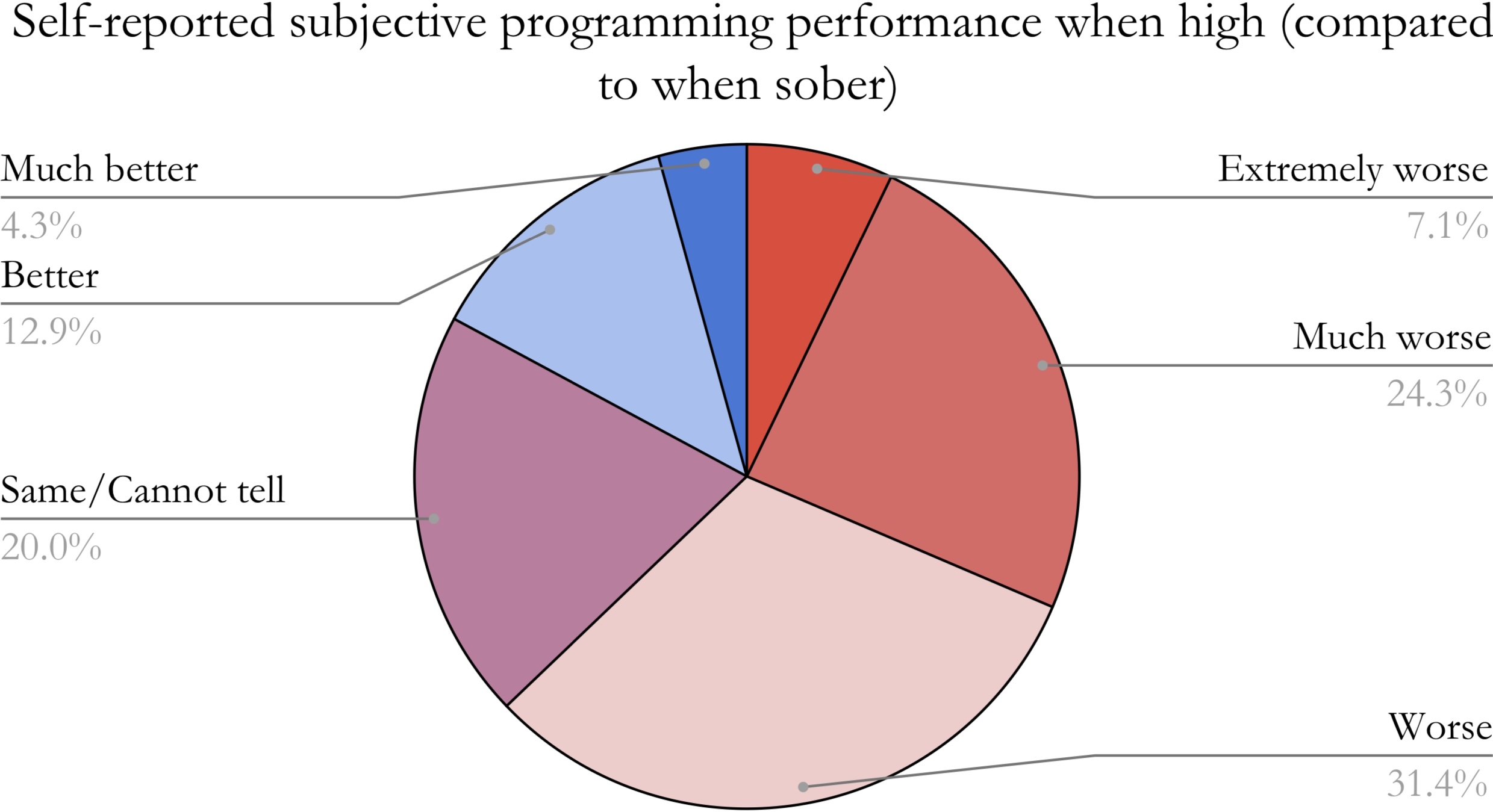}
\caption{Self-reported\label{fig-pie-chart} subjective programming performance in cannabis-using sessions compared to sober sessions. \rm{Most participants report perceiving decreased performance when high (63\%) compared to only 17\% who perceived improvement.} }
\end{figure}

For our second exploratory analysis, we investigate if participants are able to correctly perceive their relative programming performance, even when intoxicated. We explore the answer to this question because prior studies
have reported that cannabis-using developers self-assess
task contexts and potential impairment when making
usage decisions~\cite{Newman2023FromOrganizations}. To answer this question, we first investigate how participants perceived their programming performance while high vs. sober. At the end of their second sessions, we asked all participants about their performance on the ``interview style'' questions in that session compared to the previous one. 63\% thought they performed worse during their cannabis session, 20\% could not tell either way, and only 17\% thought they performed better while high. Figure~\ref{fig-pie-chart} breaks down those reports. 

While participants did, on average, have decreased performance while high, this was not universal. We thus investigate \textit{if participant perceptions of their  performance while high is accurate}: we correlate perceived performance difference (a 7-point Likert scale) with actual performance difference (average difference between percent correctness scores for all three ``interview style'' problems). 
We find a strong, positive correlation between self-reported relative performance while high and actual performance: $r=0.59$. Of the 49/75 participants who showed decreased performance on the interview problems while high, only four of them (8\%) incorrectly perceived increased performance. Together, these results indicate that programmers are generally able to accurately judge their relative programming performance, even when under the influence of cannabis.
This has implications for policy (see Section~\ref{sec:discussion}).

\begin{tcolorbox}[colback=palegreen, colframe=black, colbacktitle=white, coltitle=black, boxsep=0pt,top=6pt,bottom=6pt,left=6pt,right=6pt, beforeafter skip=6pt]
    \noindent Most programmers \textbf{can accurately judge relative programming performance while high} ($r = 0.59$). 

\end{tcolorbox}

\section{Discussion}
\label{sec:discussion}





We observe a significant impairment associated with ecologically-valid cannabis use while programming 
(10\% fewer correct tests, 10\% slower programming). At the same time, we do not observe a significant
method divergence benefit. 
Previously, anecdotal evidence (Section~\ref{sec:background}) was either conflicted or 
suggested that purported creativity benefits were worth the impairment. \textit{Our results paint a more nuanced picture, especially for situations without a robust mechanism to catch bugs or with deadline pressure.} 
We thus consider what motivates developers to use cannabis.  
After completing both sessions, we asked participants what was different when programming while high. 
The majority emphasized that it was harder to focus and easier to get distracted, 
which is contrary to prior survey results (i.e., improved focus~\cite[Tab.~3]{Endres2022HashingIt}). 
However, some participants did indicate more enjoyment, fewer worries, and decent insight into alternative perspectives. 
We note we only considered some software solution divergence aspects (e.g., we do not assess architecture or design creativity, etc.). ``Interview-style'' questions may be too structured to admit certain creative freedoms. This is relevant as programmers self-report self-regulating cannabis use by software task, using more when tasks are open-ended~\cite{Endres2022HashingIt}. Participants also reported stress from timing and researcher observation. 

Although the variance we observe in outcomes for cannabis intoxication is consistent and significant, we note that it is \textit{much less than the productivity variance already found in new hires}. 
A classic study reported 16--25$\times$ differences in coding times and 26--28$\times$ differences in 
debugging times for programmers~\cite[Tab.~III]{Sackman1968ExploratoryExperimental}, with no correlations to class grades 
or other hiring distinctions. This general pattern has continued, with 
a recent Microsoft study reporting that the time to first code check-in (in weeks), a job-relevant productivity metric, was 57\% lower in some geographic locations than others~\cite[Sec.~4.1]{Rastogi2017RampupJourney}. 
A 10\% difference is not large compared to such already-existing variance. In addition, some programmers in our sample received full correctness scores even while high, or performed better when high. Most  were 
able to accurately recognize their own cannabis-related impairment or the lack of it. Blanket employment policy may thus \emph{not} be well-motivated. 

While our study design features safeguards for the privacy and safety of our individual
participants (Section~\ref{subsec:design}), 
following Robson and McCartan~\cite[Ch.~10]{Robson2016RealWorld}, we note that job-seekers may
be considered a vulnerable group because of their economic situations. Our
research may have future implications for job-seekers (e.g., if it informs hiring policies), 
a risk the researchers, and the IRB, weighed against the benefits of conducting the research.
While we believe that the results are supportive (i.e., restrictive job policies may not be merited), we acknowledge that
the situation is nuanced. 





Anecdotally, we note that several participants reached out during the study to reschedule because they had an upcoming drug screening for a new job. This aligns with the qualitative
results of Newman \emph{et al.} who found that developers view drug policies in software as ineffective and easy to circumvent~\cite{Newman2023FromOrganizations}. Additionally, the mere existence of an anti-drug policy can serve as a deterrent for hiring and retention~\cite{Newman2023FromOrganizations}. This, combined with the low observed magnitude of cannabis impairment, may indicate that strict drug policies might not be optimal uses of resources. 

\section{Threats to Validity}

Although our observational results give confidence in our 
characterization of cannabis intoxication effects on programming, 
our results may not generalize. 
We highlight a number of considerations.

First, our participants are not a random sample of the population. Our selection may be biased
to those interested in cannabis-related studies or with a 
positive perception of cannabis. We partially mitigate this by assessing the cannabis usage history
of our participants (see Section~\ref{subsec:stimuli}). In addition, we are interested in understanding how
programmers who routinely use cannabis are affected by it in development settings: in that 
context, a participant who has not used cannabis before is less indicative of the daily impact on
a company. Similarly, the legal status of cannabis in some locations may deter participation in our
study. We partially mitigated this by recruiting in four 
US states where this sort of cannabis use
is legal. 

Second, our larger programming tasks were taken the LeetCode repository of skills-based interview
questions. These questions may not be indicative of industrial practice~\cite{Behroozi2019Hiring}. This
is partially mitigated by the fact that they are indicative of programming tasks people carry out and
study for in the hiring process. 

Third, our notions of code quality and divergence may not generalize. We assess code
correctness via tests and assess divergence and style by expert annotation. 
There are other useful notions of utility (e.g., formally proving correctness or using
other static analyses, measuring maintenance efforts, etc.) 
that we do not capture. We partially mitigate this concern, noting that
automated regression testing remains a dominant activity in SE and that 
manual assessment is relevant for both code reviewing and hiring decisions.
There are other indicators revealing creativity in software engineering problem 
solving~\cite{Groeneveld2021ExploringThe}, 
and other factors linked to programming creativity (e.g., knowledge~\cite{Hedge2014HowTo}, personality~\cite{Amin2020TheImpact, Groeneveld2021ExploringThe}), 
but we only measure divergence in products. 


Fourth, we are unable to control the amount of cannabis affecting participants. Our IRB protocol did not permit \emph{dispensing} cannabis, directing participants to take
a particular amount, or collecting blood samples --- 
instead, our \emph{observational} study involves participants using cannabis anyway. 
We partially mitigate this via photographs of cannabis products used
(and include these 
self-reported amount of marijuana and THC in our
replication materials), and by restricting attention
to one delivery method (smoking/vaping, but not edibles). 
Experienced and novice cannabis users may make different dosage decisions and have different tolerances 
(e.g.,~\cite{Boehnke2019PillsTo}), something our approach does not capture. 
\section{Conclusion}

In a controlled observational study with 74 participants, we find that at ecologically-valid dosages, \textbf{cannabis intoxication has a significant small-medium impairment on both program correctness and programming speed} ($p < 0.5, 0.22 \leq d \leq 0.44$). We did not find  evidence of cannabis increasing solution divergence. We also did not find that past cannabis usage history significantly mediates programming impairment. However, even when under the influence of cannabis, programmers correctly perceive differences in their programming performance ($r=0.59$). We hope our results contribute to the development of evidence-based policies and assist software developers in making informed decisions.


\begin{acks}
We gratefully acknowledge the partial support of the US National Science Foundation (2211749) as well as the support of Dan Clauw and Kevin Boehnke through the Michigan Psychedelic Center. 

In addition, we thank Danielle Hu for her help in running a portion of the experimental sessions, as well as multiple lab members for acting as pilot participants. We also thank Li Morrow and other members of Michigan's IRB for their help in designing an experimental protocol that prioritized participant safety and privacy.

\end{acks}

\bibliographystyle{ACM-Reference-Format}
\bibliography{endres_bib_living,endres_bib}


\begin{thebibliography}{54}


\ifx \showCODEN    \undefined \def \showCODEN     #1{\unskip}     \fi
\ifx \showDOI      \undefined \def \showDOI       #1{#1}\fi
\ifx \showISBNx    \undefined \def \showISBNx     #1{\unskip}     \fi
\ifx \showISBNxiii \undefined \def \showISBNxiii  #1{\unskip}     \fi
\ifx \showISSN     \undefined \def \showISSN      #1{\unskip}     \fi
\ifx \showLCCN     \undefined \def \showLCCN      #1{\unskip}     \fi
\ifx \shownote     \undefined \def \shownote      #1{#1}          \fi
\ifx \showarticletitle \undefined \def \showarticletitle #1{#1}   \fi
\ifx \showURL      \undefined \def \showURL       {\relax}        \fi
\providecommand\bibfield[2]{#2}
\providecommand\bibinfo[2]{#2}
\providecommand\natexlab[1]{#1}
\providecommand\showeprint[2][]{arXiv:#2}

\bibitem[Amin et~al\mbox{.}(2020)]%
        {Amin2020TheImpact}
\bibfield{author}{\bibinfo{person}{Aamir Amin}, \bibinfo{person}{Shuib Basri},
  \bibinfo{person}{Mobashar Rehman}, \bibinfo{person}{Luiz~Fernando Capretz},
  \bibinfo{person}{Rehan Akbar}, \bibinfo{person}{Abdul~Rehman Gilal}, {and}
  \bibinfo{person}{Muhammad~Farooq Shabbir}.} \bibinfo{year}{2020}\natexlab{}.
\newblock \showarticletitle{The impact of personality traits and knowledge
  collection behavior on programmer creativity}.
\newblock \bibinfo{journal}{\emph{Information and Software Technology}}
  \bibinfo{volume}{128} (\bibinfo{year}{2020}), \bibinfo{pages}{106405}.
\newblock


\bibitem[Behroozi et~al\mbox{.}(2019)]%
        {Behroozi2019Hiring}
\bibfield{author}{\bibinfo{person}{Mahnaz Behroozi}, \bibinfo{person}{Chris
  Parnin}, {and} \bibinfo{person}{Titus Barik}.}
  \bibinfo{year}{2019}\natexlab{}.
\newblock \showarticletitle{Hiring is Broken: What Do Developers Say About
  Technical Interviews?}. In \bibinfo{booktitle}{\emph{2019 {IEEE} Symposium on
  Visual Languages and Human-Centric Computing, {VL/HCC} 2019, Memphis,
  Tennessee, USA, October 14-18, 2019}},
  \bibfield{editor}{\bibinfo{person}{Justin Smith},
  \bibinfo{person}{Christopher Bogart}, \bibinfo{person}{Judith Good}, {and}
  \bibinfo{person}{Scott~D. Fleming}} (Eds.). \bibinfo{publisher}{{IEEE}
  Computer Society}, \bibinfo{pages}{1--9}.
\newblock
\urldef\tempurl%
\url{https://doi.org/10.1109/VLHCC.2019.8818836}
\showDOI{\tempurl}


\bibitem[Bennett et~al\mbox{.}(2013)]%
        {Bennett2013ComputingCreativity}
\bibfield{author}{\bibinfo{person}{Vicki~E. Bennett}, \bibinfo{person}{KyuHan
  Koh}, {and} \bibinfo{person}{Alexander Repenning}.}
  \bibinfo{year}{2013}\natexlab{}.
\newblock \showarticletitle{Computing Creativity: Divergence in Computational
  Thinking}. In \bibinfo{booktitle}{\emph{Proceeding of the 44th ACM Technical
  Symposium on Computer Science Education}}. \bibinfo{pages}{359–364}.
\newblock
\urldef\tempurl%
\url{https://doi.org/10.1145/2445196.2445302}
\showDOI{\tempurl}


\bibitem[Berman(2020)]%
        {programmingInsider}
\bibfield{author}{\bibinfo{person}{Marc Berman}.}
  \bibinfo{year}{2020}\natexlab{}.
\newblock \bibinfo{title}{How CBD Oil Can Help Programmers Focus}.
\newblock
  \bibinfo{howpublished}{\url{https://programminginsider.com/how-cbd-oil-can-help-programmers-focus/}}.
\newblock
\newblock
\shownote{Accessed: 2021-03-07}.


\bibitem[Boehnke et~al\mbox{.}(2019)]%
        {Boehnke2019PillsTo}
\bibfield{author}{\bibinfo{person}{Kevin~F. Boehnke}, \bibinfo{person}{J.~Ryan
  Scott}, \bibinfo{person}{Evangelos Litinas}, \bibinfo{person}{Suzanne
  Sisley}, \bibinfo{person}{David~A. Williams}, {and}
  \bibinfo{person}{Daniel~J. Clauw}.} \bibinfo{year}{2019}\natexlab{}.
\newblock \showarticletitle{Pills to Pot: Observational Analyses of Cannabis
  Substitution Among Medical Cannabis Users With Chronic Pain}.
\newblock \bibinfo{journal}{\emph{The Journal of Pain}} \bibinfo{volume}{20},
  \bibinfo{number}{7} (\bibinfo{year}{2019}), \bibinfo{pages}{830--841}.
\newblock
\showISSN{1526-5900}
\urldef\tempurl%
\url{https://doi.org/10.1016/j.jpain.2019.01.010}
\showDOI{\tempurl}


\bibitem[Broyd et~al\mbox{.}(2016)]%
        {Broyd2016AcuteAnd}
\bibfield{author}{\bibinfo{person}{Samantha~J. Broyd},
  \bibinfo{person}{Hendrika~H. {van Hell}}, \bibinfo{person}{Camilla Beale},
  \bibinfo{person}{Murat Yücel}, {and} \bibinfo{person}{Nadia Solowij}.}
  \bibinfo{year}{2016}\natexlab{}.
\newblock \showarticletitle{Acute and Chronic Effects of Cannabinoids on Human
  Cognition—A Systematic Review}.
\newblock \bibinfo{journal}{\emph{Biological Psychiatry}} \bibinfo{volume}{79},
  \bibinfo{number}{7} (\bibinfo{year}{2016}), \bibinfo{pages}{557--567}.
\newblock
\showISSN{0006-3223}
\urldef\tempurl%
\url{https://doi.org/10.1016/j.biopsych.2015.12.002}
\showDOI{\tempurl}
\newblock
\shownote{Cannabinoids and Psychotic Disorders}.


\bibitem[Burt et~al\mbox{.}(2021)]%
        {burt2021MechanismsOf}
\bibfield{author}{\bibinfo{person}{Thomas~S. Burt}, \bibinfo{person}{Timothy~L.
  Brown}, \bibinfo{person}{Gary Milavetz}, {and} \bibinfo{person}{Daniel~V.
  McGehee}.} \bibinfo{year}{2021}\natexlab{}.
\newblock \showarticletitle{Mechanisms of cannabis impairment: Implications for
  modeling driving performance}.
\newblock \bibinfo{journal}{\emph{Forensic Science International}}
  \bibinfo{volume}{328} (\bibinfo{year}{2021}), \bibinfo{pages}{110902}.
\newblock
\showISSN{0379-0738}
\urldef\tempurl%
\url{https://doi.org/10.1016/j.forsciint.2021.110902}
\showDOI{\tempurl}


\bibitem[Campbell(2007)]%
        {campbell2007chi}
\bibfield{author}{\bibinfo{person}{Ian Campbell}.}
  \bibinfo{year}{2007}\natexlab{}.
\newblock \showarticletitle{Chi-squared and Fisher--Irwin tests of two-by-two
  tables with small sample recommendations}.
\newblock \bibinfo{journal}{\emph{Statistics in medicine}}
  \bibinfo{volume}{26}, \bibinfo{number}{19} (\bibinfo{year}{2007}),
  \bibinfo{pages}{3661--3675}.
\newblock
\urldef\tempurl%
\url{https://doi.org/10.1002/sim.2832}
\showDOI{\tempurl}


\bibitem[Cisco(2019)]%
        {ciscoHandbook}
\bibfield{author}{\bibinfo{person}{Cisco}.} \bibinfo{year}{2019}\natexlab{}.
\newblock \bibinfo{title}{2019 Code of Business Conduct}.
\newblock
  \bibinfo{howpublished}{\url{https://www.cisco.com/c/dam/en_us/about/cobc/2019/english-2019.pdf}}.
\newblock
\newblock
\shownote{Accessed: 2021-08-09}.


\bibitem[Crawford and de~la Barra(2007)]%
        {Broderick2007EnhancingCreativity}
\bibfield{author}{\bibinfo{person}{Broderick Crawford} {and}
  \bibinfo{person}{Claudio~Le{\'o}n de~la Barra}.}
  \bibinfo{year}{2007}\natexlab{}.
\newblock \showarticletitle{Enhancing Creativity in Agile Software Teams}. In
  \bibinfo{booktitle}{\emph{Agile Processes in Software Engineering and Extreme
  Programming}}, \bibfield{editor}{\bibinfo{person}{Giulio Concas},
  \bibinfo{person}{Ernesto Damiani}, \bibinfo{person}{Marco Scotto}, {and}
  \bibinfo{person}{Giancarlo Succi}} (Eds.). \bibinfo{pages}{161--162}.
\newblock


\bibitem[Crichton et~al\mbox{.}(2021)]%
        {Crichton2021TheRole}
\bibfield{author}{\bibinfo{person}{Will Crichton}, \bibinfo{person}{Maneesh
  Agrawala}, {and} \bibinfo{person}{Pat Hanrahan}.}
  \bibinfo{year}{2021}\natexlab{}.
\newblock \showarticletitle{The Role of Working Memory in Program Tracing}. In
  \bibinfo{booktitle}{\emph{Proceedings of the 2021 CHI Conference on Human
  Factors in Computing Systems}} (Yokohama, Japan) \emph{(\bibinfo{series}{CHI
  '21})}. \bibinfo{publisher}{Association for Computing Machinery},
  \bibinfo{address}{New York, NY, USA}, Article \bibinfo{articleno}{56},
  \bibinfo{numpages}{13}~pages.
\newblock
\showISBNx{9781450380966}
\urldef\tempurl%
\url{https://doi.org/10.1145/3411764.3445257}
\showDOI{\tempurl}


\bibitem[Cuttler et~al\mbox{.}(2021)]%
        {Cuttler2021AcuteEffects}
\bibfield{author}{\bibinfo{person}{C. Cuttler}, \bibinfo{person}{E.M.
  LaFrance}, {and} \bibinfo{person}{A. Stueber}.}
  \bibinfo{year}{2021}\natexlab{}.
\newblock \showarticletitle{Acute effects of high-potency cannabis flower and
  cannabis concentrates on everyday life memory and decision making}.
\newblock \bibinfo{journal}{\emph{Sci. Rep.}} \bibinfo{volume}{11},
  \bibinfo{number}{13784} (\bibinfo{year}{2021}).
\newblock


\bibitem[Cuttler and Spradlin(2017a)]%
        {Cuttler2017MeasuringCannabis}
\bibfield{author}{\bibinfo{person}{Carrie Cuttler} {and}
  \bibinfo{person}{Alexander Spradlin}.} \bibinfo{year}{2017}\natexlab{a}.
\newblock \showarticletitle{Measuring cannabis consumption: psychometric
  properties of the daily sessions, frequency, age of onset, and quantity of
  cannabis use inventory (DFAQ-CU)}.
\newblock \bibinfo{journal}{\emph{PLoS One}} \bibinfo{volume}{12},
  \bibinfo{number}{5} (\bibinfo{year}{2017}), \bibinfo{pages}{e0178194}.
\newblock


\bibitem[Cuttler and Spradlin(2017b)]%
        {cuttler2017measuring}
\bibfield{author}{\bibinfo{person}{Carrie Cuttler} {and}
  \bibinfo{person}{Alexander Spradlin}.} \bibinfo{year}{2017}\natexlab{b}.
\newblock \showarticletitle{Measuring cannabis consumption: psychometric
  properties of the daily sessions, frequency, age of onset, and quantity of
  cannabis use inventory (DFAQ-CU)}.
\newblock \bibinfo{journal}{\emph{PLoS One}} \bibinfo{volume}{12},
  \bibinfo{number}{5} (\bibinfo{year}{2017}), \bibinfo{pages}{e0178194}.
\newblock


\bibitem[Endres et~al\mbox{.}(2022)]%
        {Endres2022HashingIt}
\bibfield{author}{\bibinfo{person}{Madeline Endres}, \bibinfo{person}{Kevin
  Boehnke}, {and} \bibinfo{person}{Westley Weimer}.}
  \bibinfo{year}{2022}\natexlab{}.
\newblock \showarticletitle{Hashing It out: A Survey of Programmers' Cannabis
  Usage, Perception, and Motivation}. In
  \bibinfo{booktitle}{\emph{International Conference on Software Engineering}}.
  \bibinfo{pages}{1107–1119}.
\newblock


\bibitem[Endres et~al\mbox{.}(2021a)]%
        {Endres2021ToRead}
\bibfield{author}{\bibinfo{person}{Madeline Endres}, \bibinfo{person}{Madison
  Fansher}, \bibinfo{person}{Priti Shah}, {and} \bibinfo{person}{Westley
  Weimer}.} \bibinfo{year}{2021}\natexlab{a}.
\newblock \showarticletitle{To read or to rotate? comparing the effects of
  technical reading training and spatial skills training on novice programming
  ability}. In \bibinfo{booktitle}{\emph{Foundations of Software Engineering}}.
  \bibinfo{pages}{754--766}.
\newblock


\bibitem[Endres et~al\mbox{.}(2021b)]%
        {Endres2021RelatingReading}
\bibfield{author}{\bibinfo{person}{Madeline Endres}, \bibinfo{person}{Zachary
  Karas}, \bibinfo{person}{Xiaosu Hu}, \bibinfo{person}{Ioulia Kovelman}, {and}
  \bibinfo{person}{Westley Weimer}.} \bibinfo{year}{2021}\natexlab{b}.
\newblock \showarticletitle{Relating Reading, Visualization, and Coding for New
  Programmers: {A} Neuroimaging Study}. In
  \bibinfo{booktitle}{\emph{International Conference on Software Engineering}}.
  \bibinfo{pages}{600--612}.
\newblock


\bibitem[Fakhoury et~al\mbox{.}(2018)]%
        {Fakhoury2018TheEffect}
\bibfield{author}{\bibinfo{person}{Sarah Fakhoury}, \bibinfo{person}{Yuzhan
  Ma}, \bibinfo{person}{Venera Arnaoudova}, {and} \bibinfo{person}{Olusola
  Adesope}.} \bibinfo{year}{2018}\natexlab{}.
\newblock \showarticletitle{The Effect of Poor Source Code Lexicon and
  Readability on Developers' Cognitive Load}. In
  \bibinfo{booktitle}{\emph{International Conference on Program
  Comprehension}}.
\newblock


\bibitem[Ford et~al\mbox{.}(2022)]%
        {Ford2022ATale}
\bibfield{author}{\bibinfo{person}{Denae Ford},
  \bibinfo{person}{Margaret{-}Anne~D. Storey}, \bibinfo{person}{Thomas
  Zimmermann}, \bibinfo{person}{Christian Bird}, \bibinfo{person}{Sonia Jaffe},
  \bibinfo{person}{Chandra~Shekhar Maddila}, \bibinfo{person}{Jenna~L. Butler},
  \bibinfo{person}{Brian Houck}, {and} \bibinfo{person}{Nachiappan Nagappan}.}
  \bibinfo{year}{2022}\natexlab{}.
\newblock \showarticletitle{A Tale of Two Cities: Software Developers Working
  from Home during the {COVID-19} Pandemic}.
\newblock \bibinfo{journal}{\emph{{ACM} Trans. Softw. Eng. Methodol.}}
  \bibinfo{volume}{31}, \bibinfo{number}{2} (\bibinfo{year}{2022}),
  \bibinfo{pages}{27:1--27:37}.
\newblock
\urldef\tempurl%
\url{https://doi.org/10.1145/3487567}
\showDOI{\tempurl}


\bibitem[Fridberg et~al\mbox{.}(2010)]%
        {Fridberg2010CognitiveMechanisms}
\bibfield{author}{\bibinfo{person}{Daniel~J. Fridberg}, \bibinfo{person}{Sarah
  Queller}, \bibinfo{person}{Woo-Young Ahn}, \bibinfo{person}{Woojae Kim},
  \bibinfo{person}{Anthony~J. Bishara}, \bibinfo{person}{Jerome~R. Busemeyer},
  \bibinfo{person}{Linda Porrino}, {and} \bibinfo{person}{Julie~C. Stout}.}
  \bibinfo{year}{2010}\natexlab{}.
\newblock \showarticletitle{Cognitive mechanisms underlying risky
  decision-making in chronic cannabis users}.
\newblock \bibinfo{journal}{\emph{Journal of Mathematical Psychology}}
  \bibinfo{volume}{54}, \bibinfo{number}{1} (\bibinfo{year}{2010}),
  \bibinfo{pages}{28--38}.
\newblock
\showISSN{0022-2496}
\urldef\tempurl%
\url{https://doi.org/10.1016/j.jmp.2009.10.002}
\showDOI{\tempurl}
\newblock
\shownote{Contributions of Mathematical Psychology to Clinical Science and
  Assessment}.


\bibitem[Fry et~al\mbox{.}(2012)]%
        {Fry2012AHuman}
\bibfield{author}{\bibinfo{person}{Zachary~P. Fry}, \bibinfo{person}{Bryan
  Landau}, {and} \bibinfo{person}{Westley Weimer}.}
  \bibinfo{year}{2012}\natexlab{}.
\newblock \showarticletitle{A human study of patch maintainability}. In
  \bibinfo{booktitle}{\emph{{ISSTA}}}. \bibinfo{publisher}{{ACM}},
  \bibinfo{pages}{177--187}.
\newblock


\bibitem[Gonzales and Cunninghham(2015)]%
        {Gonzales2015ThePromise}
\bibfield{author}{\bibinfo{person}{Joseph~E. Gonzales} {and}
  \bibinfo{person}{Corbin~A. Cunninghham}.} \bibinfo{year}{2015}\natexlab{}.
\newblock \showarticletitle{The promise of pre registration in psychological
  research}.
\newblock \bibinfo{journal}{\emph{American Psychological Association}}
  (\bibinfo{year}{2015}).
\newblock


\bibitem[Graziotin et~al\mbox{.}(2014)]%
        {Graziotin2014HappySolve}
\bibfield{author}{\bibinfo{person}{Daniel Graziotin}, \bibinfo{person}{Xiaofeng
  Wang}, {and} \bibinfo{person}{Pekka Abrahamsson}.}
  \bibinfo{year}{2014}\natexlab{}.
\newblock \showarticletitle{Happy software developers solve problems better:
  psychological measurements in empirical software engineering}.
\newblock \bibinfo{journal}{\emph{PeerJ}}  \bibinfo{volume}{2}
  (\bibinfo{date}{March} \bibinfo{year}{2014}), \bibinfo{pages}{e289}.
\newblock
\showISSN{2167-8359}


\bibitem[Groeneveld et~al\mbox{.}(2021)]%
        {Groeneveld2021ExploringThe}
\bibfield{author}{\bibinfo{person}{Wouter Groeneveld}, \bibinfo{person}{Laurens
  Luyten}, \bibinfo{person}{Joost Vennekens}, {and} \bibinfo{person}{Kris
  Aerts}.} \bibinfo{year}{2021}\natexlab{}.
\newblock \showarticletitle{Exploring the Role of Creativity in Software
  Engineering}. In \bibinfo{booktitle}{\emph{43rd {IEEE/ACM} International
  Conference on Software Engineering: Software Engineering in Society, {ICSE}
  {(SEIS)} 2021, May 25-28, 2021}}. \bibinfo{publisher}{{IEEE}},
  \bibinfo{address}{Madrid, Spain}, \bibinfo{pages}{1--9}.
\newblock
\urldef\tempurl%
\url{https://doi.org/10.1109/ICSE-SEIS52602.2021.00009}
\showDOI{\tempurl}


\bibitem[Hegde and Walia(2014)]%
        {Hedge2014HowTo}
\bibfield{author}{\bibinfo{person}{Reshma Hegde} {and}
  \bibinfo{person}{Gursimran Walia}.} \bibinfo{year}{2014}\natexlab{}.
\newblock \showarticletitle{How to enhance the creativity of software
  developers: A systematic literature review}.
\newblock \bibinfo{journal}{\emph{International Conference on Software
  Engineering and Knowledge Engineering}} (\bibinfo{year}{2014}),
  \bibinfo{pages}{229--234}.
\newblock


\bibitem[IBM(2018)]%
        {ibmHandbook}
\bibfield{author}{\bibinfo{person}{IBM}.} \bibinfo{year}{2018}\natexlab{}.
\newblock \bibinfo{title}{Business Conduct Guidelines}.
\newblock
  \bibinfo{howpublished}{\url{https://www.ibm.com/investor/att/pdf/BCG_accessible_2019.pdf}}.
\newblock
\newblock
\shownote{Accessed: 2021-08-09}.


\bibitem[Kelion(2014)]%
        {bbcFBI2014}
\bibfield{author}{\bibinfo{person}{Leo Kelion}.}
  \bibinfo{year}{2014}\natexlab{}.
\newblock \bibinfo{title}{FBI 'could hire hackers on cannabis' to fight
  cybercrime}.
\newblock
  \bibinfo{howpublished}{\url{https://www.bbc.com/news/technology-27499595}}.
\newblock
\newblock
\shownote{Accessed: 2021-03-07}.


\bibitem[Kowal et~al\mbox{.}(2015)]%
        {Kowal2015CannabisAnd}
\bibfield{author}{\bibinfo{person}{Mikael~A Kowal}, \bibinfo{person}{Arno
  Hazekamp}, \bibinfo{person}{Lorenza~S Colzato}, \bibinfo{person}{Henk van
  Steenbergen}, \bibinfo{person}{Nic~JA van~der Wee}, \bibinfo{person}{Jeffrey
  Durieux}, \bibinfo{person}{Meriem Manai}, {and} \bibinfo{person}{Bernhard
  Hommel}.} \bibinfo{year}{2015}\natexlab{}.
\newblock \showarticletitle{Cannabis and creativity: highly potent cannabis
  impairs divergent thinking in regular cannabis users}.
\newblock \bibinfo{journal}{\emph{Psychopharmacology}} \bibinfo{volume}{232},
  \bibinfo{number}{6} (\bibinfo{year}{2015}), \bibinfo{pages}{1123--1134}.
\newblock


\bibitem[Kroon et~al\mbox{.}(2021)]%
        {Kroon2021TheShort}
\bibfield{author}{\bibinfo{person}{Emese Kroon}, \bibinfo{person}{Lauren
  Kuhns}, {and} \bibinfo{person}{Janna Cousijn}.}
  \bibinfo{year}{2021}\natexlab{}.
\newblock \showarticletitle{The short-term and long-term effects of cannabis on
  cognition: recent advances in the field}.
\newblock \bibinfo{journal}{\emph{Current Opinion in Psychology}}
  \bibinfo{volume}{38} (\bibinfo{year}{2021}), \bibinfo{pages}{49--55}.
\newblock
\showISSN{2352-250X}
\urldef\tempurl%
\url{https://doi.org/10.1016/j.copsyc.2020.07.005}
\showDOI{\tempurl}
\newblock
\shownote{Cannabis}.


\bibitem[Krueger et~al\mbox{.}(2020)]%
        {Krueger2020NeurologicalDivide}
\bibfield{author}{\bibinfo{person}{Ryan Krueger}, \bibinfo{person}{Yu Huang},
  \bibinfo{person}{Xinyu Liu}, \bibinfo{person}{Tyler Santander},
  \bibinfo{person}{Westley Weimer}, {and} \bibinfo{person}{Kevin Leach}.}
  \bibinfo{year}{2020}\natexlab{}.
\newblock \showarticletitle{Neurological Divide: An fMRI Study of Prose and
  Code Writing}. In \bibinfo{booktitle}{\emph{International Conference on
  Software Engineering}}.
\newblock


\bibitem[LaFrance and Cuttler(2017)]%
        {LaFrance2017InspiredBy}
\bibfield{author}{\bibinfo{person}{Emily~M. LaFrance} {and}
  \bibinfo{person}{Carrie Cuttler}.} \bibinfo{year}{2017}\natexlab{}.
\newblock \showarticletitle{Inspired by Mary Jane? Mechanisms underlying
  enhanced creativity in cannabis users}.
\newblock \bibinfo{journal}{\emph{Consciousness and Cognition}}
  \bibinfo{volume}{56} (\bibinfo{year}{2017}), \bibinfo{pages}{68--76}.
\newblock
\showISSN{1053-8100}
\urldef\tempurl%
\url{https://doi.org/10.1016/j.concog.2017.10.009}
\showDOI{\tempurl}


\bibitem[Markoff(2005)]%
        {Markoff2005WhatThe}
\bibfield{author}{\bibinfo{person}{John Markoff}.}
  \bibinfo{year}{2005}\natexlab{}.
\newblock \bibinfo{booktitle}{\emph{What the dormouse said: How the sixties
  counterculture shaped the personal computer industry}}.
\newblock \bibinfo{publisher}{Penguin Group}, \bibinfo{address}{New York, NY,
  USA}.
\newblock


\bibitem[McDowell(2015)]%
        {McDowell2015Craking}
\bibfield{author}{\bibinfo{person}{Gayle~Laakmann McDowell}.}
  \bibinfo{year}{2015}\natexlab{}.
\newblock \bibinfo{booktitle}{\emph{Cracking the coding interview—189
  programming questions and solutions.}}
\newblock \bibinfo{publisher}{CareerCup}.
\newblock


\bibitem[Mohanani et~al\mbox{.}(2017)]%
        {Mohanani2017Perceptions}
\bibfield{author}{\bibinfo{person}{Rahul Mohanani}, \bibinfo{person}{Prabhat
  Ram}, \bibinfo{person}{Ahmed Lasisi}, \bibinfo{person}{Paul Ralph}, {and}
  \bibinfo{person}{Burak Turhan}.} \bibinfo{year}{2017}\natexlab{}.
\newblock \showarticletitle{Perceptions of Creativity in Software Engineering
  Research and Practice}. In \bibinfo{booktitle}{\emph{2017 43rd Euromicro
  Conference on Software Engineering and Advanced Applications (SEAA)}}.
  \bibinfo{pages}{210--217}.
\newblock
\urldef\tempurl%
\url{https://doi.org/10.1109/SEAA.2017.21}
\showDOI{\tempurl}


\bibitem[Newman et~al\mbox{.}(2023)]%
        {Newman2023FromOrganizations}
\bibfield{author}{\bibinfo{person}{Kaia Newman}, \bibinfo{person}{Madeline
  Endres}, \bibinfo{person}{Westley Weimer}, {and} \bibinfo{person}{Brittany
  Johnson}.} \bibinfo{year}{2023}\natexlab{}.
\newblock \showarticletitle{From Organizations to Individuals: Psychoactive
  Substance Use By Professional Programmers}. In
  \bibinfo{booktitle}{\emph{International Conference on Software Engineering}}.
  \bibinfo{pages}{665--677}.
\newblock


\bibitem[Nguyen and Shanks(2009)]%
        {Nguyen2009AFramework}
\bibfield{author}{\bibinfo{person}{Lemai Nguyen} {and} \bibinfo{person}{Graeme
  Shanks}.} \bibinfo{year}{2009}\natexlab{}.
\newblock \showarticletitle{A framework for understanding creativity in
  requirements engineering}.
\newblock \bibinfo{journal}{\emph{Information and Software Technology}}
  \bibinfo{volume}{51}, \bibinfo{number}{3} (\bibinfo{year}{2009}),
  \bibinfo{pages}{655--662}.
\newblock
\showISSN{0950-5849}


\bibitem[Norman(2010)]%
        {likertStatistics}
\bibfield{author}{\bibinfo{person}{Geoffrey Norman}.}
  \bibinfo{year}{2010}\natexlab{}.
\newblock \showarticletitle{Likert scales, levels of measurement and the
  “laws” of statistics}.
\newblock \bibinfo{journal}{\emph{Advances in health sciences education :
  theory and practice}} \bibinfo{volume}{15}, \bibinfo{number}{5}
  (\bibinfo{date}{02} \bibinfo{year}{2010}), \bibinfo{pages}{625--32}.
\newblock
\urldef\tempurl%
\url{https://doi.org/10.1007/s10459-010-9222-y}
\showDOI{\tempurl}


\bibitem[Peitek et~al\mbox{.}(2021)]%
        {Peitek2021ProgramComprehension}
\bibfield{author}{\bibinfo{person}{Norman Peitek}, \bibinfo{person}{Sven Apel},
  \bibinfo{person}{Chris Parnin}, \bibinfo{person}{André Brechmann}, {and}
  \bibinfo{person}{Janet Siegmund}.} \bibinfo{year}{2021}\natexlab{}.
\newblock \showarticletitle{Program Comprehension and Code Complexity Metrics:
  An fMRI Study}. In \bibinfo{booktitle}{\emph{2021 IEEE/ACM 43rd International
  Conference on Software Engineering (ICSE)}}. \bibinfo{pages}{524--536}.
\newblock
\urldef\tempurl%
\url{https://doi.org/10.1109/ICSE43902.2021.00056}
\showDOI{\tempurl}


\bibitem[Rastogi et~al\mbox{.}(2017)]%
        {Rastogi2017RampupJourney}
\bibfield{author}{\bibinfo{person}{Ayushi Rastogi}, \bibinfo{person}{Suresh
  Thummalapenta}, \bibinfo{person}{Thomas Zimmermann},
  \bibinfo{person}{Nachiappan Nagappan}, {and} \bibinfo{person}{Jacek
  Czerwonka}.} \bibinfo{year}{2017}\natexlab{}.
\newblock \showarticletitle{Ramp-up Journey of New Hires: Do Strategic
  Practices of Software Companies Influence Productivity?}. In
  \bibinfo{booktitle}{\emph{Proceedings of the 10th Innovations in Software
  Engineering Conference}}. \bibinfo{pages}{107–111}.
\newblock


\bibitem[Robson and McCartan(2016)]%
        {Robson2016RealWorld}
\bibfield{author}{\bibinfo{person}{Colin Robson} {and} \bibinfo{person}{Kieran
  McCartan}.} \bibinfo{year}{2016}\natexlab{}.
\newblock \bibinfo{booktitle}{\emph{Real world research: a resource for users
  of social research methods in applied settings}}.
\newblock \bibinfo{publisher}{Wiley}.
\newblock


\bibitem[Rogeberg and Elvik(2016)]%
        {Rogeberg2016Cannabis}
\bibfield{author}{\bibinfo{person}{Ole Rogeberg} {and} \bibinfo{person}{Rune
  Elvik}.} \bibinfo{year}{2016}\natexlab{}.
\newblock \showarticletitle{The effects of cannabis intoxication on motor
  vehicle collision revisited and revised}.
\newblock \bibinfo{journal}{\emph{Addiction}} \bibinfo{volume}{111},
  \bibinfo{number}{8} (\bibinfo{year}{2016}), \bibinfo{pages}{1348--1359}.
\newblock
\urldef\tempurl%
\url{https://doi.org/10.1111/add.13347}
\showDOI{\tempurl}
\showeprint{https://onlinelibrary.wiley.com/doi/pdf/10.1111/add.13347}


\bibitem[Runco and Okuda(1988)]%
        {Runco1988Problem}
\bibfield{author}{\bibinfo{person}{M.~A. Runco} {and} \bibinfo{person}{S.M.
  Okuda}.} \bibinfo{year}{1988}\natexlab{}.
\newblock \showarticletitle{Problem discovery, divergent thinking, and the
  creative process}.
\newblock \bibinfo{journal}{\emph{Journal of Youth and Adolescence}}
  \bibinfo{volume}{17}, \bibinfo{number}{3} (\bibinfo{date}{06}
  \bibinfo{year}{1988}), \bibinfo{pages}{211--220}.
\newblock
\urldef\tempurl%
\url{https://doi.org/10.1007/BF01538162}
\showDOI{\tempurl}


\bibitem[Sackman et~al\mbox{.}(1968)]%
        {Sackman1968ExploratoryExperimental}
\bibfield{author}{\bibinfo{person}{H. Sackman}, \bibinfo{person}{W.~J.
  Erikson}, {and} \bibinfo{person}{E.~E. Grant}.}
  \bibinfo{year}{1968}\natexlab{}.
\newblock \showarticletitle{Exploratory Experimental Studies Comparing Online
  and Offline Programming Performance}.
\newblock \bibinfo{journal}{\emph{Commun. ACM}} \bibinfo{volume}{11},
  \bibinfo{number}{1} (\bibinfo{date}{jan} \bibinfo{year}{1968}),
  \bibinfo{pages}{3–11}.
\newblock
\showISSN{0001-0782}


\bibitem[Seabold and Perktold(2010)]%
        {seabold2010statsmodels}
\bibfield{author}{\bibinfo{person}{Skipper Seabold} {and}
  \bibinfo{person}{Josef Perktold}.} \bibinfo{year}{2010}\natexlab{}.
\newblock \showarticletitle{statsmodels: Econometric and statistical modeling
  with python}. In \bibinfo{booktitle}{\emph{9th Python in Science
  Conference}}. \bibinfo{publisher}{SciPy}, \bibinfo{address}{Austin, TX, US},
  \bibinfo{pages}{92--96}.
\newblock


\bibitem[Shihab et~al\mbox{.}(2023)]%
        {msr2023}
\bibfield{author}{\bibinfo{person}{Emad Shihab}, \bibinfo{person}{Patanamon
  Thongtanunam}, {and} \bibinfo{person}{Bogdan Vasilescu}.}
  \bibinfo{year}{2023}\natexlab{}.
\newblock \showarticletitle{Mining Software Repositories}.
\newblock \bibinfo{journal}{\emph{{IEEE}}} (\bibinfo{year}{2023}).
\newblock


\bibitem[Siegmund et~al\mbox{.}(2014)]%
        {Siegmund2014UnderstandingUnderstanding}
\bibfield{author}{\bibinfo{person}{Janet Siegmund}, \bibinfo{person}{Christian
  K{\"a}stner}, \bibinfo{person}{Sven Apel}, \bibinfo{person}{Chris Parnin},
  \bibinfo{person}{Anja Bethmann}, \bibinfo{person}{Thomas Leich},
  \bibinfo{person}{Gunter Saake}, {and} \bibinfo{person}{Andr{\'e} Brechmann}.}
  \bibinfo{year}{2014}\natexlab{}.
\newblock \showarticletitle{Understanding understanding source code with
  functional magnetic resonance imaging}. In
  \bibinfo{booktitle}{\emph{Proceedings of the 36th International Conference on
  Software Engineering}}. \bibinfo{pages}{378--389}.
\newblock


\bibitem[Simmons et~al\mbox{.}(2020)]%
        {Simmons2020PreregistrationWhy}
\bibfield{author}{\bibinfo{person}{Joseph~P. Simmons}, \bibinfo{person}{Leif~D.
  Nelson}, {and} \bibinfo{person}{Uri Simonsohn}.}
  \bibinfo{year}{2020}\natexlab{}.
\newblock \showarticletitle{Pre-registration: Why and How}.
\newblock \bibinfo{journal}{\emph{J. Society for Consumer Psychology}}
  (\bibinfo{date}{Dec.} \bibinfo{year}{2020}).
\newblock
\urldef\tempurl%
\url{https://doi.org/10.1002/jcpy.1208}
\showDOI{\tempurl}


\bibitem[Team(2020)]%
        {UNPressReport}
\bibfield{author}{\bibinfo{person}{United Nations~Press Team}.}
  \bibinfo{year}{2020}\natexlab{}.
\newblock \bibinfo{booktitle}{\emph{UNODC World Drug Report 2020: Global drug
  use rising; while COVID-19 has far reaching impact on global drug markets}}.
\newblock United Nations.
\newblock
\urldef\tempurl%
\url{https://www.unodc.org/unodc/press/releases/2020/June/media-advisory---global-launch-of-the-2020-world-drug-report.html}
\showURL{%
\tempurl}


\bibitem[Virtanen et~al\mbox{.}(2020)]%
        {2020SciPy-NMeth}
\bibfield{author}{\bibinfo{person}{Pauli Virtanen}, \bibinfo{person}{Ralf
  Gommers}, \bibinfo{person}{Travis~E. Oliphant}, \bibinfo{person}{Matt
  Haberland}, \bibinfo{person}{Tyler Reddy}, \bibinfo{person}{David
  Cournapeau}, \bibinfo{person}{Evgeni Burovski}, \bibinfo{person}{Pearu
  Peterson}, \bibinfo{person}{Warren Weckesser}, \bibinfo{person}{Jonathan
  Bright}, \bibinfo{person}{St{\'e}fan~J. {van der Walt}},
  \bibinfo{person}{Matthew Brett}, \bibinfo{person}{Joshua Wilson},
  \bibinfo{person}{K.~Jarrod Millman}, \bibinfo{person}{Nikolay Mayorov},
  \bibinfo{person}{Andrew R.~J. Nelson}, \bibinfo{person}{Eric Jones},
  \bibinfo{person}{Robert Kern}, \bibinfo{person}{Eric Larson},
  \bibinfo{person}{C~J Carey}, \bibinfo{person}{{\.I}lhan Polat},
  \bibinfo{person}{Yu Feng}, \bibinfo{person}{Eric~W. Moore},
  \bibinfo{person}{Jake {VanderPlas}}, \bibinfo{person}{Denis Laxalde},
  \bibinfo{person}{Josef Perktold}, \bibinfo{person}{Robert Cimrman},
  \bibinfo{person}{Ian Henriksen}, \bibinfo{person}{E.~A. Quintero},
  \bibinfo{person}{Charles~R. Harris}, \bibinfo{person}{Anne~M. Archibald},
  \bibinfo{person}{Ant{\^o}nio~H. Ribeiro}, \bibinfo{person}{Fabian Pedregosa},
  \bibinfo{person}{Paul {van Mulbregt}}, {and} \bibinfo{person}{{SciPy 1.0
  Contributors}}.} \bibinfo{year}{2020}\natexlab{}.
\newblock \showarticletitle{{{SciPy} 1.0: Fundamental Algorithms for Scientific
  Computing in Python}}.
\newblock \bibinfo{journal}{\emph{Nature Methods}}  \bibinfo{volume}{17}
  (\bibinfo{year}{2020}), \bibinfo{pages}{261--272}.
\newblock
\urldef\tempurl%
\url{https://doi.org/10.1038/s41592-019-0686-2}
\showDOI{\tempurl}


\bibitem[Walsh(2011)]%
        {Walsh2011DrugsThe}
\bibfield{author}{\bibinfo{person}{Charlotte Walsh}.}
  \bibinfo{year}{2011}\natexlab{}.
\newblock \showarticletitle{Drugs, the Internet and change}.
\newblock \bibinfo{journal}{\emph{Journal of psychoactive drugs}}
  \bibinfo{volume}{43}, \bibinfo{number}{1} (\bibinfo{year}{2011}),
  \bibinfo{pages}{55--63}.
\newblock


\bibitem[Walton(2019)]%
        {simpleProgrammer}
\bibfield{author}{\bibinfo{person}{Mary Walton}.}
  \bibinfo{year}{2019}\natexlab{}.
\newblock \bibinfo{title}{Programming and Cannabis — 5 Things to Know}.
\newblock
  \bibinfo{howpublished}{\url{https://simpleprogrammer.com/programming-and-cannabis/}}.
\newblock
\newblock
\shownote{Accessed: 2021-03-07}.


\bibitem[{W}es {M}c{K}inney(2010)]%
        {pandas}
\bibfield{author}{\bibinfo{person}{{W}es {M}c{K}inney}.}
  \bibinfo{year}{2010}\natexlab{}.
\newblock \showarticletitle{{D}ata {S}tructures for {S}tatistical {C}omputing
  in {P}ython}. In \bibinfo{booktitle}{\emph{{P}roceedings of the 9th {P}ython
  in {S}cience {C}onference}}, \bibfield{editor}{\bibinfo{person}{{S}t\'efan
  van~der {W}alt} {and} \bibinfo{person}{{J}arrod {M}illman}} (Eds.).
  \bibinfo{publisher}{SciPy}, \bibinfo{address}{Austin, TX, US},
  \bibinfo{pages}{56--61}.
\newblock
\urldef\tempurl%
\url{https://doi.org/10.25080/Majora-92bf1922-00a}
\showDOI{\tempurl}


\bibitem[Wood(2021)]%
        {businessWire}
\bibfield{author}{\bibinfo{person}{Laura Wood}.}
  \bibinfo{year}{2021}\natexlab{}.
\newblock \bibinfo{booktitle}{\emph{Global Cannabis Market (2020 to 2026) -
  Emergence of Cannabis Legalization in Asia-Pacific Presents Opportunities -
  ResearchAndMarkets.com}}.
\newblock Business Wire.
\newblock
\urldef\tempurl%
\url{https://www.businesswire.com/news/home/20210216005966/en/Global-Cannabis-Market-2020-to-2026---Emergence-of-Cannabis-Legalization-in-Asia-Pacific-Presents-Opportunities---ResearchAndMarkets.com/}
\showURL{%
\tempurl}


\bibitem[Yamada(2018)]%
        {Yamada2018HowTo}
\bibfield{author}{\bibinfo{person}{Yuki Yamada}.}
  \bibinfo{year}{2018}\natexlab{}.
\newblock \showarticletitle{How to Crack Pre-registration: Toward Transparent
  and Open Science}.
\newblock \bibinfo{journal}{\emph{Frontiers in Psychology}}
  \bibinfo{volume}{9}, \bibinfo{number}{1831} (\bibinfo{year}{2018}).
\newblock
\urldef\tempurl%
\url{https://doi.org/10.3389/fpsyg.2018.01831}
\showDOI{\tempurl}


\end{thebibliography}

\end{document}